\documentclass[10pt,letterpaper]{article}
\usepackage[top=0.85in,left=1.2in,footskip=0.75in]{geometry}

\usepackage{amsmath,amssymb}

\usepackage{changepage}

\usepackage[utf8x]{inputenc}

\usepackage{textcomp,marvosym}

\usepackage{cite}

\usepackage{nameref,hyperref}

\usepackage[right]{lineno}

\usepackage{microtype}
\DisableLigatures[f]{encoding = *, family = * }

\usepackage[table]{xcolor}

\usepackage{array}

\usepackage{subfig}

\usepackage{multirow}

\usepackage{cases, enumitem}

\usepackage{rotating}

\newcolumntype{+}{!{\vrule width 2pt}}

\newlength\savedwidth

\usepackage[aboveskip=1pt,labelfont=bf,labelsep=period,justification=raggedright,singlelinecheck=off]{caption}

\makeatletter
\renewcommand{\@biblabel}[1]{\quad#1.}
\makeatother

\usepackage{lastpage,fancyhdr,graphicx}
\usepackage{epstopdf}
\fancyheadoffset[L]{2.25in}
\fancyfootoffset[L]{2.25in}
\begin{document}
\vspace*{0.2in}

\begin{flushleft}
{\Large
\textbf\newline{Bayesian interpolation for power laws in neural data analysis} 
}
\newline
\\
Iván A. Davidovich, 
Yasser Roudi
\\
\bigskip
Kavli Institute for Systems Neuroscience, NTNU, Trondheim, Norway
\\
\bigskip

ivan.a.davidovich@ntnu.no, yasser.roudi@ntnu.no

\end{flushleft}
\section*{Abstract}
Power laws arise in a variety of phenomena ranging from matter undergoing phase transition to the distribution of word frequencies in the English language. Usually, their presence is only apparent when data is abundant, and accurately determining their exponents often requires even larger amounts of data. As the scale of recordings in neuroscience becomes larger, an increasing number of studies attempt to characterise potential power-law relationships in neural data. In this paper, we aim to discuss the potential pitfalls that one faces in such efforts and to promote a Bayesian interpolation framework for this purpose. We apply this framework to synthetic data and to data from a recent study of large-scale recordings in mouse primary visual cortex (V1), where the exponent of a power-law scaling in the data played an important role: its value was argued to determine whether the population's stimulus-response relationship is smooth, and experimental data was provided to confirm that this is indeed so. Our analysis shows that with such data types and sizes as we consider here, the best-fit values found for the parameters of the power law and the uncertainty for these estimates are heavily dependent on the noise model assumed for the estimation, the range of the data chosen, and (with all other things being equal) the particular recordings. It is thus challenging to offer a reliable statement about the exponents of the power law. Our analysis, however, shows that this does not affect the conclusions regarding the smoothness of the population response to low-dimensional stimuli but casts doubt on those to natural images. We discuss the implications of this result for the neural code in the V1 and offer the approach discussed here as a framework that future studies, perhaps exploring larger ranges of data, can employ as their starting point to examine power-law scalings in neural data.


\section{Introduction}\label{sec-intro}

A power law refers to the situation in which two variables $x$ and $y$ are related to each other as $y= b x^{\alpha}$. They are interesting and important for a number of reasons. One reason comes primarily from the physics of critical phenomena when systems go through a phase transition: in some phase transitions, where a system goes from one phase to another, power laws describe the relationship between some of the quantities that characterise the state of the system \cite{goldenfeld2018lectures,simons1997phase}. A typical example of this is that of some materials such as metallic alloys, where as temperature, $T$, is lowered, a spontaneous transition from being non-magnetised to magnetised occurs. This happens at a critical temperature, $T_c$, and for $(T_c-T)/T= \epsilon>0$, one has a power-law relationship, called a scaling law, $m \sim \epsilon^{\alpha}$ for the mean magnetisation of the alloy, $m$. The power, or exponent, $\alpha$ in this case is called a critical exponent. Similarly, in self-organised criticality, namely where dynamical systems self-tune their parameters to stay near the boundary between different phases, power-law scalings are abundant. In fact, the original paper on self-organised criticality was an attempt to explain the $1/f$ noise \cite{bak1987self}. The existence of scaling laws and the precise value of the relevant exponents in such critical systems are considered pillars of critical phenomena and supported by a large body of both theoretical and experimental work \cite{stanley1999scaling,kadanoff1967static}.

Another reason why power laws are important is that they can be useful for describing the probability distribution of certain stochastic events. The Central Limit Theorem (CLT) states that the sum, $s$, of a large number of finite-variance stochastic variables is distributed according to a Gaussian distribution. However, if the variance is not finite, the probability distribution of the sum converges to heavy-tailed, non-Gaussian distributions \cite{feller1957introduction}. For these so called {\em sum-stable distributions}, $p(s)$, the probability of $s$, behaves as a power law for large $s$ \cite{feller1957introduction}. These distributions play an important role in the study of a plethora of disparate systems and phenomena, from the distribution of the magnitudes of earthquakes \cite{ishimoto1936observations} to the number of occurrences of words in a text \cite{piantadosi2014zipf}. It is known that studying these distributions requires a certain degree of care so as not to draw misguided conclusions, for instance by considering alternative probability models as a potentially better description of the dataset, \textit{i.e.} performing model selection \cite{clauset2009power,newman2005power,chapman2002extremum}.

In neuroscience, power laws have been used both for describing scaling laws \cite{amit1985storing,treves1991determines,katkov2015effects, stringer2019high} and as probability distributions describing certain events, e.g. the size of clusters of simultaneously spiking neurons, in both theoretical models and experiments \cite{beggs2003neuronal,beggs2012being}. Our focus here is mostly on studies that describe scaling laws. For instance, theoretical models of memory predict power-law scalings relating the capacity of neural networks to the number of modifiable connections per neuron \cite{amit1985storing,treves1991determines}; the pre-factor and exponent may well depend on details of the model and can in many cases be precisely calculated \cite{sompolinsky1986neural, treves1991determines, roudi2007balanced}. In some other studies, the starting point is not a theoretical mechanistic model, but an experimental dataset that appears to exhibit a scaling law \cite{miller2009power,pozzorini2013temporal}. Some studies lend a significant amount of weight, not only to the existence of a scaling law in the dataset, but also to the specific values obtained for its parameters~\cite{richardson2002comparing,simkin2014stochastic,stringer2019high}. The rationale behind this would be, for instance, that competing hypotheses all predict scaling laws but with different powers. The range of data utilised in different studies that analyse or report power laws in neuroscience varies widely, in some cases being as small as one or two decades \cite{shriki2013neuronal}. This does not mean that those studies necessarily miss something, as this would depend on why the presence of the power law and the precise value of its parameters are important. For instance, in the physics of critical phenomena the precise value of the exponents in a power law are important for checking the validity of theoretically predicted relationships between critical exponents. The violation of the predicted relationships may have far-reaching implications, e.g. in the case of the so-called hyper-scaling relationships, those that depend on the dimensionality of the model. There is thus considerable work on estimating the relevant exponents and checking the validity of hyper-scaling relationships in different systems \cite{fisher1967theory,stell1971relation,simons1997phase,schwartz1991breakdown,campbell2019hyperscaling}. 

In a recent publication, Stringer et al~\cite{stringer2019high} argue through a theoretical approach that, in order for the population neural response to vary smoothly with the stimulus, the eigenvalue corresponding to the $n$-th principal component of the neural population response to a $d$-dimensional stimulus should scale as $n^{-\alpha}$ with $\alpha \ge \alpha_d=1+1/d$. In other words, they argue that the precise value of the exponent $\alpha$ is quite important, as it determines if the neural code is smooth or not, depending on whether it is above or below the theoretically prescribed lower bound. Large scale calcium imaging recordings of the mouse visual cortex in response to a range of stimuli with different dimensionalities, from natural scene stimuli presumed to be very high dimensional to low-dimensional gratings, are presented to conclude that the theoretical bound on $\alpha$ required for this smooth stimulus-response relationship is indeed satisfied. But in this and other cases, how certain can we be about the results obtained from analysing experimental data? The importance of answering this question can be appreciated if we note that, for instance, in Stringer et al~\cite{stringer2019high}, in the case of natural scene stimuli, exponents from individual recordings were both above and below the critical value. However, the authors substantiated their conclusion relying on the average of the estimated exponents, which turned out to be slightly above the critical value. Performing the same procedure, the average exponent for the lower-dimensional stimuli yielded a value much larger than the critical lower bound. Is this difference between natural scene and low-dimensional stimuli important? Is it appropriate to average the estimated exponents? How can uncertainties about the estimates be quantified and to what extent do they affect the conclusions?

Answering these questions requires first establishing that a power law is indeed the best description of the data, as opposed to other descriptions, namely other functional relationships between $x$ and $y$. Assuming the power law to be indeed present, the second, related, problem is to estimate its parameters: the pre-factor, $b$, and often most importantly, the exponent $\alpha$. Solving both problems in a statistically sound way may be quite cumbersome. The presence of noise in the data and not having data spanning a sufficient range can lead to misleading results: identifying something that is not a power law as a power law, or finding incorrect values for the exponent or the pre-factor. These issues are not commonly addressed when discussing power-law relationships in neural data even though, as we hope becomes clear in this paper and as has already been noted in other areas of science, the implications could be far-reaching \cite{kadanoff1967static,clauset2009power,stumpf2012critical}. 
 
Here we focus on the second problem mentioned above, that of parameter estimation; that is, when the presence of a power-law scaling is already assumed. We study this issue using data we generate synthetically from known distributions and also real data from the aforementioned study of~\cite{stringer2019high}. The authors there estimated the exponent of the presumed power law by performing a linear regression, over a preselected range of data, in log-log space. In this paper, instead, we perform a more nuanced analysis by recasting the problem of inferring the best-fit parameters in terms of the framework of Bayesian interpolation~\cite{MacKay92a}. Stating the problem in this framework forces us to make our assumptions explicit, thus allowing us to evaluate their impact on the results in a clear and systematic way. It also allows us to find not only the best-fit parameters, but also the uncertainty in our estimates. Finally, we can also evaluate the idea of pooling estimates from different recordings, which is, and will likely continue to be, important in neuroscience. We consider three different models, differing in the way we model noise, and consider different ranges of both the synthetic and real data. 

Our results on the synthetic data show a number of clear patterns regarding how the estimates of the parameters of a known power law behave when there are mismatches between the generative model and the interpolation model, or when the range of data is either insufficient or inappropriately chosen. These patterns can be used as a guide and first step checks when analysing real data. When this is done on the neural responses to natural stimuli referred to above, we find that the differences produced by the different models and assumptions, as well as the uncertainties on the exponent estimated with each model, are too large to permit a conclusive statement as to whether or not the exponent is above the theoretical lower bound. On the other hand, when neural responses to low-dimensional stimuli are studied, despite the fact that variations are still seen on the estimated exponent, in all cases the exponent is far above the predicted lower bound; one can thus say, with much more confidence, that the lower bound in not violated in this case. We will discuss the implications of this observation on the processing of visual stimuli in the primary visual cortex (V1) of mice.

As noted above, here we focus on estimating the parameters of the power law and will not address the problem of whether the data is actually best described by a power law or if other functional forms are more appropriate. However, we note that the best-fit parameters estimation explored here is a necessary step for such a model comparison~\cite{MacKay92a}, if one is to proceed without any \textit{ad hoc} additions.

\section{Materials and methods}

\subsection*{Bayesian Interpolation}\label{sec-bay-int}

\subsubsection*{Model description}\label{sec-model-desc}

The general framework that we employ here is that of Bayesian interpolation~\cite{MacKay92a} applied and adapted to the case where we want to infer parameters that are not simply the coefficients in a linear combination of functional forms (see also~\cite{bretthorst1990bayesian}). Specifically, given a dataset of ordered pairs of the form $D = \{(x_i,d_i), i\in \mathbb{N}, i\leq N\}$, we will consider $3$ different noise models for our data: 

\begin{samepage}
\begin{enumerate}[label=\alph*.]
\item {\bf Power-Linear model (PL)}: Additive Gaussian noise of variance $\beta^{-1}$ on the data as presented, 
\item {\bf Straight-Line model (SL)}: Additive Gaussian noise of variance $\beta^{-1}$ on the logarithm of the data,  
\item {\bf Weighted Straight-Line model (WSL)}: Additive Gaussian noise of variance $x_i \beta^{-1}$ on the logarithm of the data. 
\end{enumerate}
\end{samepage}
In what follows, we define each of these models in turn.

The PL model is perhaps the one that most immediately comes to mind when thinking of fitting a power law to our dataset; we assume that the data contains some additive Gaussian noise and try to interpolate them with a power-law functional form as they are. Given the dataset $D$ described above, the PL model is then defined by:
\begin{subequations}
\begin{numcases}{}
	&$d_i = y(x_i) + \nu_i, \quad \forall i\geq {\rm min}$\\
	&$y(x_i) = b \left(\frac{x_i}{x_{\rm min}}\right)^{-\alpha}$  \label{eq-plaw} \\
	&$\nu_i \sim \mathcal{N}\left(0,\frac{1}{\beta}\right)$\label{eq-noisemodel}
\end{numcases}
\end{subequations}

The parametrization of the power law in Eq.~(\ref{eq-plaw}) is useful during the numerical exploration of the parameter space, but note that it is of course equivalent to the more usual form $a x_i^{-\alpha}$ via $b = a x_{\rm min}^{-\alpha}$. As shown in Eq. (\ref{eq-noisemodel}), we are assuming an additive Gaussian noise model here, which simplifies the calculations. The cut-off parameter $x_{\rm min}$ is included to allow for the common situation in which we don't want to fit a power-law functional form to our whole dataset but rather only for large enough values of the independent variable $x$; the reason for this is that typically in real data the relationship between $d$ and $x$ does not follow a supposed power law for small values of $x$. In principle, one can also include an $x_{\rm max}$ parameter to add an upper cut-off to the power-law behaviour; here we have chosen instead to simply explore numerically the effects of using the whole dataset or the specific value of $x_{\rm max}$ used by the original authors of the experimental dataset we analyse \cite{stringer2019high}. We also explore the effect of changing $x_{\rm min}$ and compare our results to theirs.

The SL model is born out of the popularity of linear regression on the log-log plot of the data as a way of performing inference on the parameters of an assumed power law. Defining $w_i := \log_{10} x_i$ and $f_i := \log_{10} d_i$, $\forall i$, the SL model is defined by:
\begin{subequations}
\begin{numcases}{}
	&$f_i = z(w_i) + \nu_i, \quad \forall i\geq {\rm min}$ \label{eq-SLmodel} \\
	&$z(w_i) = \log_{10} b + \alpha \left(w_{\rm min} - w_i\right)$ \label{eq-sline}\\
	&$\nu_i \sim \mathcal{N}\left(0,\frac{1}{\beta}\right)$ \label{eq-noise_2}
\end{numcases}
\end{subequations}

Note that once again we're assuming additive Gaussian noise, but this time on the logarithm of the values, as is done implicitly when minimizing squared errors in a linear regression. This corresponds to a multiplicative log-normal noise in our original variables. Consequently, it will not necessarily be the case that the results of our inference procedure in this scenario match the ones in our first scenario, when using the data in its original form.

Finally, the WSL model comes from the need to compare our results with those obtained by the authors of~\cite{stringer2019high}. There, a linear interpolation in the log-log plot of the data was performed, as in the SL model. However, examination of the code used in that article reveals that the fitting of parameters was performed using a weighted version of least-squares which is equivalent to the assumption that the variance of the Gaussian noise in~(\ref{eq-noise_2}) is proportional to $x_i$. Therefore, the WSL model is also defined by Eqs.~(\ref{eq-SLmodel}-\ref{eq-noise_2}) but replacing $\beta^{-1}$ by $x_i \beta^{-1}$ in Eq.(\ref{eq-noise_2}).

In each of our models there are $3$ parameters to be determined: $b$, $\alpha$ and $x_{\rm min}$. Ideally we would determine all of them via Bayesian inference. However, determining an estimate of $x_{\rm min}$ is particularly challenging, which is why we will assume it given and, as mentioned above, explore numerically how our results change as we vary this parameter. There is also an auxiliary parameter that describes the noise variance, $\beta$, which, as we will show below, we will infer from the data.

\subsubsection*{Posterior density, likelihood and parameter determination}\label{sec-post_setup}

If we assume $x_{\rm min}$ to be given, for each of the chosen models, $\mathcal{A} \in \{{\rm PL, SL, WSL}\}$, we would like to determine $P_{\mathcal{A}}(b,\alpha|D)$, that is, the posterior probability density for the model parameters $b$ and $\alpha$ given the data, $D$, and our choice of priors over model parameters (which we omit from the notation for simplicity). In practice, we will obtain this density by considering instead $P_{\mathcal{A}}(b,\alpha|D,\beta)$ and integrating over the nuisance parameter $\beta$, which represents the inverse of the noise variance.

To start, we note that for all three models described in the previous section, we can write the likelihood given a dataset $D = \{(x_i,d_i), i\in \mathbb{N}, i\leq N\}$ as
\begin{align}
	P_{\mathcal{A}}(D|b,\alpha,\beta) = \frac{\exp\left[-\beta \, E_{\mathcal{A}}(D|b,\alpha)\right]}{Z_{\mathcal{A}}(\beta)},
\end{align}
where the subscripts denote the models ($\mathcal{A} \in \{{\rm PL, SL, WSL}\}$) and the energy functions, $E_{\mathcal{A}}$, take the form
\begin{subequations}
\begin{align}
	&E_{\rm PL} \equiv \frac{1}{2}\sum_{i = {\rm min}}^{N} \left[y(x_i) - d_i\right]^2\\
	&E_{\rm SL} \equiv \frac{1}{2}\sum_{i = {\rm min}}^{N} \left[z(w_i) - f_i\right]^2, \\  
	&E_{\rm WSL} \equiv \frac{1}{2}\sum_{i = {\rm min}}^{N} \frac{\left[z(w_i) - f_i\right]^2}{x_i}
\end{align}
\label{E-eq}
\end{subequations}
and the normalizations are
\begin{subequations}
\begin{align}
	&Z_{\rm PL/SL} \equiv \left(\frac{2 \pi}{\beta}\right)^{\frac{N-{\rm min}}{2}}\\
	&Z_{\rm WSL} \equiv \prod_{i = {\rm min}}^{N} \left(\frac{2 \pi x_i}{\beta}\right)^{\tfrac{1}{2}}.
\end{align}
\end{subequations}

\noindent The simplest way of finding the best-fit values for $\alpha$ and $b$, in the absence of strong priors, is to use maximum likelihood. Since the normalization, $Z_{\mathcal{A}}(\beta)$, does not depend on the values of $b$ and $\alpha$, the maximum likelihood estimates of these parameters are simply those that maximize $E_{\mathcal{A}}$, and do not require knowledge of $\beta$. 

To determine the uncertainty of the estimated best-fit values of $\alpha$ and $b$, we calculate the posterior density over the parameters
\begin{align}\label{eq-post_marginal}
	P_{\mathcal{A}}(b,\alpha|D) = \int d\beta P_{\mathcal{A}}(b,\alpha|D,\beta) \, P_{\mathcal{A}}(\beta|D).
\end{align}
In general, this is a difficult integral to perform but it turns out that a reasonable estimate can be obtained as~\cite{MacKay92a}

\begin{equation}
 P_{\mathcal{A}}(b,\alpha|D) =  P_{\mathcal{A}}(b,\alpha|D,\hat{\beta})
\end{equation}
where here $\hat{\beta}$ is found as the value that maximizes the evidence $P_{\mathcal{A}}(D|\beta)$, that is
\begin{equation}
\hat{\beta} = {\rm argmax}_{\beta} P_{\mathcal{A}}(D|\beta)
\end{equation}
Calculating the evidence, in turn, requires us to calculate the integral 
\begin{equation}
P_{\mathcal{A}}(D|\beta) = \int db\ d \alpha \ P_{\mathcal{A}}(D|b,\alpha, \beta) P(b,\alpha)
\end{equation}
where $P(b,\alpha)$ is the prior over the parameters $b$ and $\alpha$. Calculating this integral is also difficult, but again, a result can be obtained by using a Gaussian approximation of the integrand~\cite{MacKay92a}. This yields
\begin{align}\label{eq-logevidence}
	\ln P_{\mathcal{A}}\left(D|\beta\right) = - \beta E_{\mathcal{A}}^{\rm ML} - \frac{1}{2} \ln \det \beta B - \ln Z_{\mathcal{A}}(\beta) + \ln \frac{2\pi}{(b_M-b_0)(\alpha_M-\alpha_0)} ,
\end{align} 
where $B = \nabla^2 E_{\mathcal{A}}^{\rm ML}$, calculated using the relevant expression in  Eq. \eqref{E-eq} at the maximum-likelihood values $\hat{\alpha}$ and $\hat{b}$. In writing Eq.~\eqref{eq-logevidence} we have also assumed a uniform prior over $\alpha$ and $b$ in the predefined ranges $[b_0, b_M]$ and $[\alpha_0,\alpha_M]$ as

\begin{equation}\label{eq-flat_prior}
        P(b,\alpha)= 
\begin{cases}
    [(b_M-b_0)(\alpha_M-\alpha_0)]^{-1} &, \text{if } b_0 \leq b \leq b_M, \alpha_0 \leq \alpha \leq \alpha_M\\
    0              &, \text{otherwise}.
\end{cases}
\end{equation}

The expressions for the entries of $B$ for the WSL model can be calculated to be

\begin{subequations}
\begin{numcases}{}
	&$B^{\rm WSL}_{11} = \frac{\partial^2 E_{\rm WSL}}{\partial^2 b} = \frac{1}{b^2 \ln 10} \sum_{i = {\rm min}}^{N} \frac{1}{x_i}\left[\frac{1}{\ln 10} - \left(\log_{10} b + \alpha (w_{\rm min} - w_i) - f_i \right) \right]$, \\
	&$B^{\rm WSL}_{12} = \frac{\partial^2 E_{\rm WSL}}{\partial \alpha \partial b} = \frac{\partial^2 E_{\rm WSL}}{\partial b \partial \alpha} = B^{\rm WSL}_{21} = \frac{1}{b \ln 10}\sum_{i = {\rm min}}^{N} \frac{(w_{\rm {\rm min}} - w_i)}{x_i}$, \\  
	&$B^{\rm WSL}_{22} = \frac{\partial^2 E_{\rm WSL}}{\partial^2 \alpha} = \sum_{i = min}^{N} \frac{(w_{\rm min} - w_i)^2}{x_i}$.
\end{numcases}
\end{subequations}
The corresponding matrix for the SL model is obtained by simply removing the $(x_i)^{-1}$ factors in each of the elements above.

We can then find the value of $\hat{\beta}$ that maximizes the evidence for the SL and WSL models analytically from Eq.~(\ref{eq-logevidence}). This yields
\begin{align}\label{eq-noise_est_flat}
	\hat{\beta} = \frac{\left((N-x_{\rm min}) - 2\right)}{2 E^{\rm ML}_{\mathcal{A}}}.
\end{align}  
With this value of $\beta$ we can estimate the posterior density for $b$ and $\alpha$, which with flat priors will amount simply to a rescaling of the likelihood.

For the PL model, Eq.~(\ref{eq-logevidence}) is only a second-order approximation (in $b$ and $\alpha$; see \cite{MacKay92a}), with the $B$ matrix given by:

\begin{subequations}
\begin{numcases}{}
	&$B^{\rm PL}_{11} = \sum_{i = {\rm min}}^{N} \left(\frac{x_i}{x_{\rm min}}\right)^{-2\alpha}$, \\
	&$B^{\rm PL}_{12} =  B^{\rm PL}_{21} = -\sum_{i = min}^{N} \left[2 b \left(\frac{x_i}{x_{\rm min}}\right)^{-\alpha} - d_i \right] \left(\frac{x_i}{x_{\rm min}}\right)^{-\alpha} \ln\left(\frac{x_i}{x_{\rm min}}\right)$, \\  
	&$B^{\rm PL}_{22} = \sum_{i = min}^{N} \left[2 b \left(\frac{x_i}{x_{\rm min}}\right)^{-\alpha} - d_i \right] b \left(\frac{x_i}{x_{\rm min}}\right)^{-\alpha} \ln^2\left(\frac{x_i}{x_{\rm min}}\right)$.
\end{numcases}
\end{subequations}

 Numerical integration of the evidence without analytical approximations is in principle possible for this model, but in practical terms this is unfeasible for the kind of datasets we want to analyse, as we quickly run into numbers that are either too large or too small for our computer systems to represent accurately. Nevertheless, as we will see in the \nameref{sec-results} section, numerical tests on synthetic data show that Eq.~(\ref{eq-noise_est_flat}) provides an estimation for the order of magnitude of $\beta$ in the PL model which is on par with those obtained for the SL and WSL models. We can then use this value of $\beta$ together with the $B^{\rm PL}$ matrix to estimate the uncertainty around our best-fit values of $b$ and $\alpha$ for the PL model.

\subsection*{The data}

The experimental data that we use here as an example are taken from~\cite{stringer2019high}. This data is organized in different sets of ordered pairs, each corresponding to ranked eigenvalues (or variances) in the neural responses of the visual cortex of awake mice to different sets of stimuli, as measured through calcium imaging. The details of the experimental procedures and pre-processing of data that lead to the ordered pairs of ranks and eigenvalues can be found in the original publication. We will not be concerned here with how the data were recorded or processed in order to obtain the ranked eigenvalues; instead we focus on the interpolation problem. 

\section{Results}\label{sec-results}
\subsection*{Posterior density and its maxima: synthetic data}\label{sec-pl-mock}

Before applying the framework presented in the \nameref{sec-bay-int} section to the experimental dataset, we first apply it to synthetic data to address a number of questions when we know the ground truth. Specifically, we would like to study (a) how using a noise model for inference on samples generated from another noise model affects the results, (b) how changing $x_{\rm min}$ and the cut-off influences the estimates (next section) (c) how much variability there is in the estimated parameters from one sample to another when both are generated from the same noise model, and how estimates from these different samples can be combined together. This latter issue is addressed because in neural data one often deals with different recordings that are presumed to reflect a single underlying law; for example, in~\cite{stringer2019high} the mean of the exponents estimated from individual recordings was taken as reflecting a common true exponent of the underlying power law.

To this end, we first generated data according to each of our $3$ noise models: SL, WSL and PL. We chose the values $\alpha^{\rm tr} = 1.04$, $b^{\rm tr} = 0.01$ and $x^{\rm tr}_{\rm min} = 11$ for a more straightforward comparison to  \cite{stringer2019high}; the superscript ${\rm tr}$ simply indicates that these are the true values for the parameters used when generating these datasets. For $x < x^{\rm tr}_{\rm min}$ we use the functional form $b^{\rm tr} \sqrt{x/x^{\rm tr}_{\rm min}}$. When it comes to generating the noise contribution for each model, we also employ values of the respective $\beta$ which are of the same order of magnitude as those which we will later show to be present in the experimental data (see Table~\ref{tab-betas}); we use $\beta^{-\frac{1}{2}}_{\rm SL} = 10^{-2}$, $\beta^{-\frac{1}{2}}_{\rm WSL} = 10^{-3}$ and $\beta^{-\frac{1}{2}}_{\rm PL} = 10^{-4}$.

\begin{figure}[ph!]
	\centering
	\includegraphics[width=0.7\paperwidth]{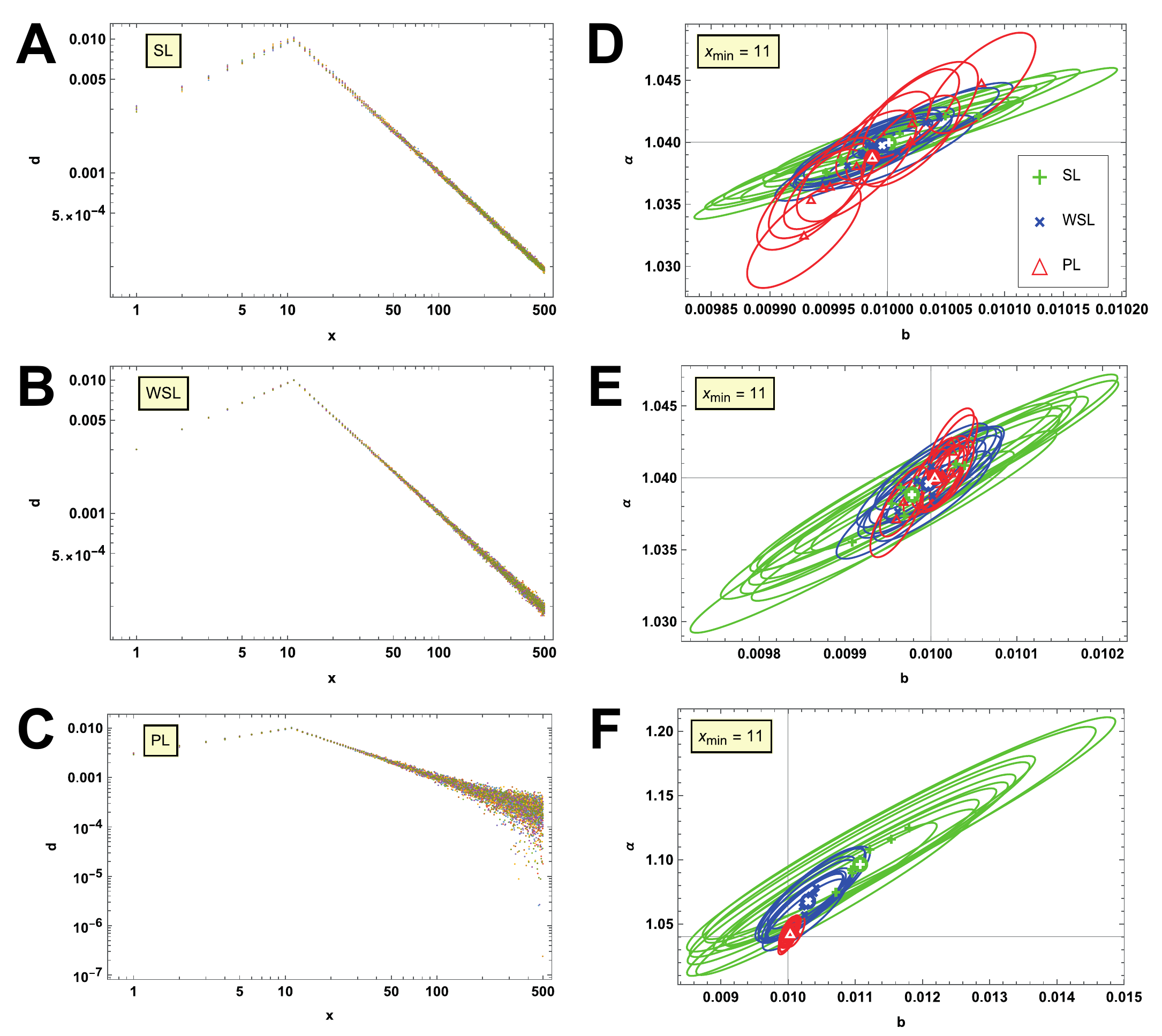}
	\caption{\textbf{Inference results on synthetic data.} On the left column \textbf{(A-C)} we present $10$ datasets per panel, each generated according to one of our $3$ noise models, as indicated in the main text. Each colour corresponds to a different dataset. On the right column \textbf{(D-F)}, we summarize the results of performing inference on the datasets of the corresponding left panel with all $3$ models; the one that generated the data and the ones that didn't. We present this information in the form of the position of the maximum of the posterior distribution for $b$ and $\alpha$ as well as the $99\%$ confidence region ellipse around that maximum. To aid in the discussion, coloured disks with white symbols indicate the position of the average of the posterior maxima for the corresponding inference model, over the $10$ datasets. Black lines indicate the true values of the parameters.}
	\label{fig_synth_data}
	\end{figure}

A sample of the datasets generated in this way is shown on the left column of Fig.~\ref{fig_synth_data} for the different noise models. The results of estimating $b$ and $\alpha$ on these samples, using the noise model which generated the data as well as the other two noise models are summarized on the right column of Fig.~\ref{fig_synth_data}. There, we show the position of the maximum of the posterior distribution (which is the same as the maximum-likelihood estimate in our case) for $b$ and $\alpha$. The corresponding $99\%$ confidence regions for each inference model and for all $10$ datasets are also shown. These uncertainty ellipses were obtained by calculating the covariance matrix for our parameters at the position of the maximum, i.e. the inverse of the Hessian matrix $\beta B$ in Eq.~(\ref{eq-logevidence}). 

When the model used for inference is consistent with the model that generated the data, it is clear from Fig.~\ref{fig_synth_data}D-F that the $99\%$ confidence region of the posterior, for each sample, is centered near the true parameter values. When the model is inconsistent with the generative model, this is no longer the case. Fig.~\ref{fig_synth_data}D shows that when the data is generated from SL but estimation is done through PL, the $99\%$ ellipses often do not include the true parameter values. The peak of the posterior seems, however, to be above or below the true values with equal chance depending on the particular sample. When PL is used on data generated from WSL, the situation is similar (Fig.~\ref{fig_synth_data}E). On the other hand, when SL or WSL are applied to data from PL (Fig.~\ref{fig_synth_data}F), overestimated values for both $b$ and $\alpha$ are more likely in each of the different samples. Although we notice that the inconsistent models often assign a high posterior to the true parameter values, the patterns observed in this figure will become important when we analyse the real data.    

\begin{figure}[ph!]
	\centering
	\includegraphics[width=0.7\paperwidth]{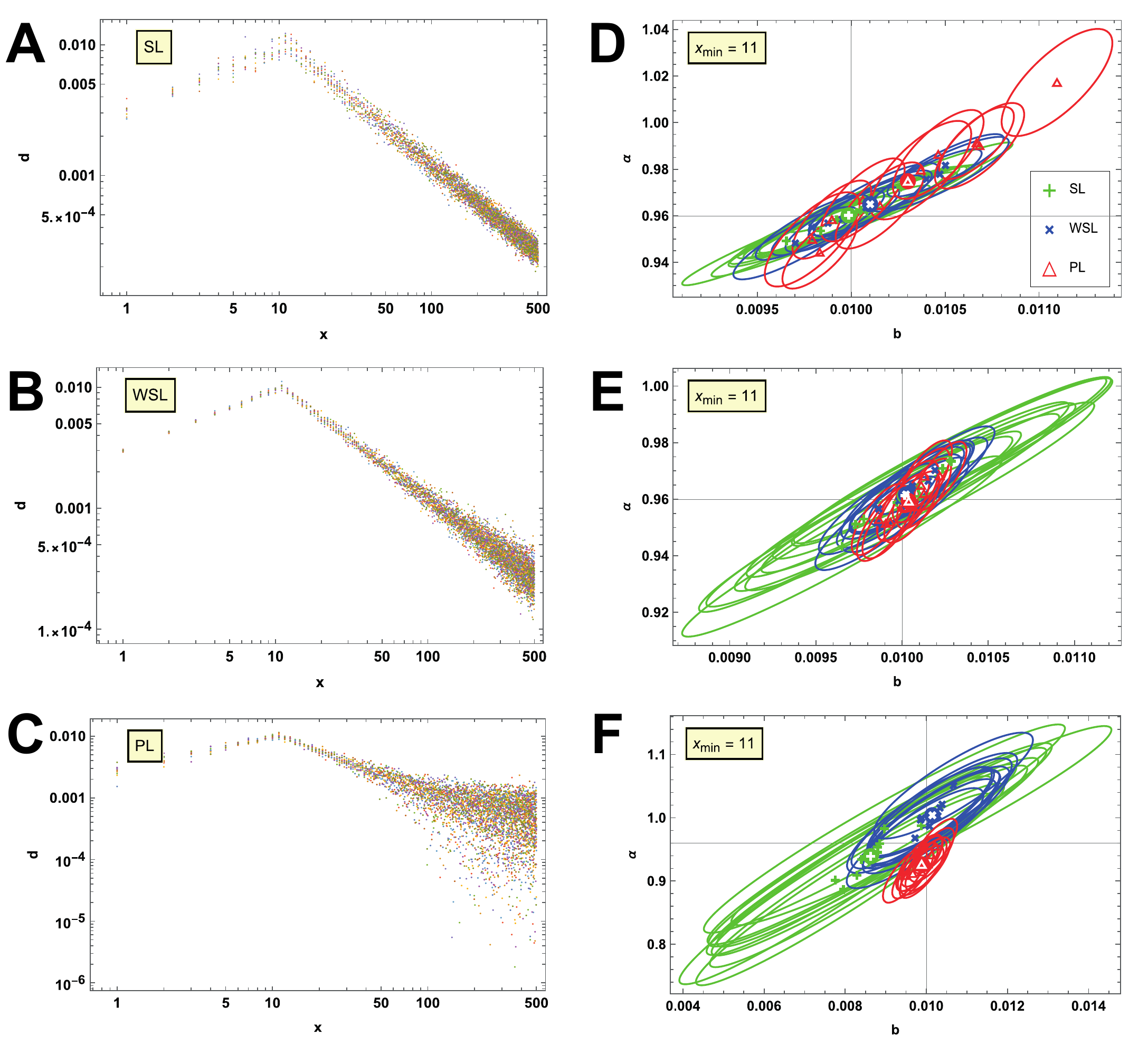}
	\caption{\textbf{Inference results on synthetic data with larger noise variance and lower $\alpha$.} On the left column \textbf{(A-C)} we present $10$ datasets per panel, each generated according to one of our $3$ noise models, as indicated in the main text. Compared to Fig.~\ref{fig_synth_data}, the data presented here was created using a noise variance that is $25$ times larger and a lower value of the power-law exponent, $\alpha^{\rm tr} = 0.96$. Each colour corresponds to a different dataset. On the right column \textbf{(D-F)}, we summarize the results of performing inference on the datasets of the corresponding left panel with all $3$ models; the one that generated the data and the ones that didn't. We present this information in the form of the position of the maximum of the posterior distribution for $b$ and $\alpha$ as well as the $99\%$ confidence region ellipse around that maximum. To aid in the discussion, coloured disks with white symbols indicate the position of the average of the posterior maxima for the corresponding inference model, over the $10$ datasets. Black lines indicate the true values of the parameters. For the reasons explained in the text, the datasets used in \textbf{(F)} differ from those of \textbf{(C)} in the sign of some of the points with a large value of $x$.}
	\label{fig_synth_datax5}
\end{figure}

As mentioned in the~\nameref{sec-intro}, one of the central points in~\cite{stringer2019high} is to determine whether the exponent of an assumed power law in their main dataset is larger or smaller than $1$. The results of Fig.~\ref{fig_synth_data} and the discussion above suggest that a mismatch between generative and inference noise models could cause a true value of the exponent, $\alpha^{\rm tr}$, which is smaller than $1$ to be estimated as being larger than $1$ instead. We explore this possibility by generating new synthetic data, now with $\alpha^{\rm tr} = 0.96$. At the same time, we also explore how sensitive our results are to $\beta^{-1}$, that is, to the noise variance used to generate the data. In Fig.~\ref{fig_synth_datax5} we present the datasets and inference results obtained with $\alpha^{\rm tr} = 0.96$ and values of $\beta^{-1/2}$ which are $5$ times larger than the ones used for generating Fig.~\ref{fig_synth_data}. Additionally, Fig.~\ref{fig_SI_mockx10} in the \nameref{sec-supp} shows the same kind of results but now with values of $\beta^{-1/2}$ which are $10$ times larger than those of Fig.~\ref{fig_synth_data}. Figs.~\ref{fig_synth_datax5} and ~\ref{fig_SI_mockx10} demonstrate examples for which one gets estimates of $\alpha > 1$ while $\alpha^{\rm tr} < 1$ (panels D and F). These figures also highlight the importance of quantifying the amount of uncertainty in our inference (the curvature of the posterior around its maximum), since we can see that in most cases where the best estimate for $\alpha$ is above $1$, values below $1$ are also compatible with our data within the $99\%$ confidence region; we also see a few examples of the opposite situation. The information contained in the posterior density would indeed allow us to calculate the probability of $\alpha$ being above or below this threshold.

We note that the PL model can generate negative values, which causes issues with estimation using either SL or WSL since they involve taking a logarithm of those values. For this reason, to produce the plots in the F panels, the inferences were run on the absolute values of the data from the PL model.
In Fig.~\ref{fig_synth_datax5} there were around $70$ points in each dataset which changed sign when taking the absolute value, while this happened for $\sim120$ points in Fig.~\ref{fig_synth_datax5} and only around $2$ points in Fig.~\ref{fig_synth_data}; this is the reason why the PL model performs well for inference in Fig.~\ref{fig_synth_data}F but shows a mismatch in the other two figures. This mismatch would not be present if we remove the absolute value, leaving the data unmodified. While the data modified by taking the absolute value could plausibly have been obtained from the PL model, this is unlikely. It is perhaps more useful to interpret the results in Fig.~\ref{fig_synth_datax5}F and Fig.~\ref{fig_SI_mockx10}F as corresponding to inference on data generated with an unspecified fourth model.

In Figs.~\ref{fig_synth_data}, \ref{fig_synth_datax5} and Fig.~\ref{fig_SI_mockx10} we show only $10$ samples from each generative model, which is similar to the number of recordings in the experimental dataset we turn to in the next section. The situation is the same when more samples are considered.

\begin{figure}[h]
	\includegraphics[width=0.8\paperwidth]{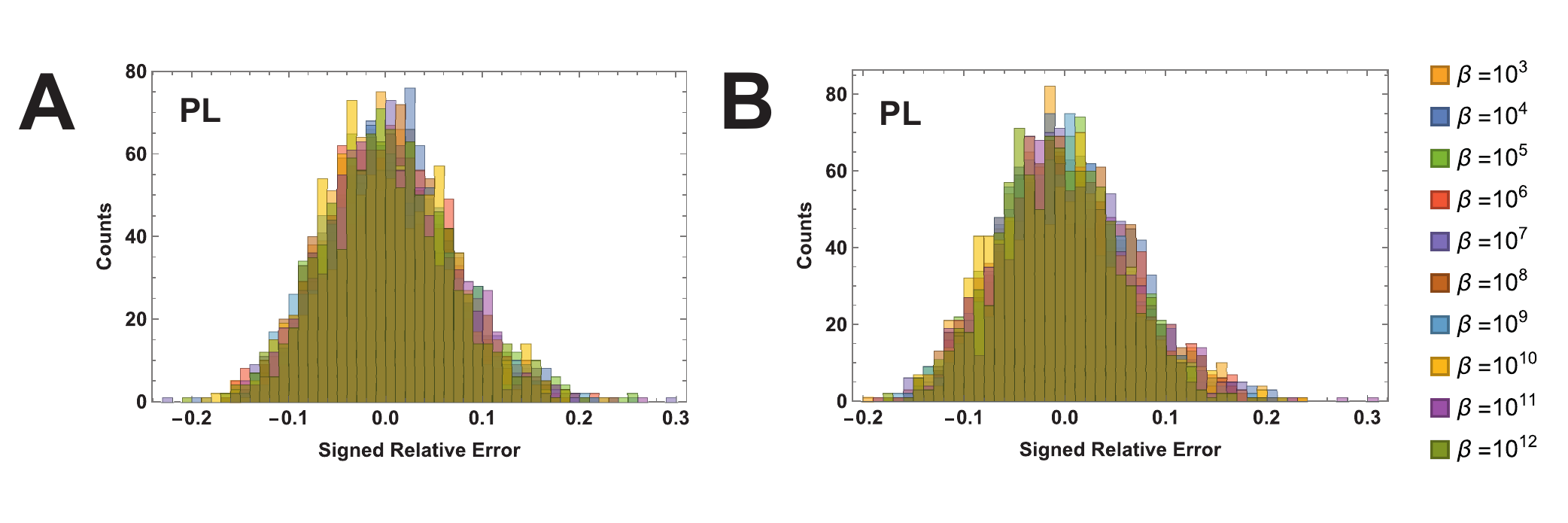}
	\caption{\textbf{Performance of the $\beta$ estimation with the PL model.} \textbf{(A)} Histograms of the signed relative error in the estimation of $\beta$ when using Eq.~(\ref{eq-noise_est_flat}), after estimating $b$ and $\alpha$ with the PL model on corresponding synthetic datasets, and for a range of ground truth values of $\beta$ as indicated in the legend. Each histogram was built from $1000$ datasets per ground truth value of $\beta$. \textbf{(B)} The same as in \textbf{(A)} but now the ground truth value of $b$ and $\alpha$ has been employed when using Eq.~(\ref{eq-noise_est_flat}). Bin widths are $0.01$ for all panels.}
	\label{fig-beta_scan}
\end{figure}

As discussed earlier, in the Bayesian inference we estimate the value of the parameter controlling the noise variance, $\beta$, for each of our models by using Eq.~(\ref{eq-noise_est_flat}). We test the quality of this approximation by, once again, generating synthetic data with the same ground truth parameters used for Fig.~\ref{fig_synth_data}, but now varying the value of $\beta$ in the range $10^3$ to $10^{12}$. We add noise according to each of our models and then use them to infer the best-fit values of $b$ and $\alpha$, which we use in turn to estimate $\beta$ with Eq.~(\ref{eq-noise_est_flat}). Fig.~\ref{fig-beta_scan}A shows histograms for the PL model of the signed relative error in the $\beta$ estimation (\textit{i.e.} $(\beta - \beta^{\rm tr})/\beta^{\rm tr})$). The corresponding histograms for the SL and WSL models are qualitatively very similar to these and can be found in Fig~\ref{fig_SI_beta}A. Each histogram contains $1000$ datasets per ground truth value of $\beta$. We see from these histograms that the relative error only rarely exceeds the $20\%$ level in either direction, suggesting that Eq.~(\ref{eq-noise_est_flat}) provides a good order-of-magnitude estimation for $\beta$ for all $3$ models within the wide range of noise variances we studied. It's worth noting that our calculations include not only the estimation of $\beta$ but also of $b$ and $\alpha$, which could be dominating the deviations we observe. To disentangle these two contributions, we remove this element by directly using the ground truth values $b^{\rm tr}$ and $\alpha^{\rm tr}$ in Eq.~(\ref{eq-noise_est_flat}). These results are shown in Figs.~\ref{fig-beta_scan}B and~\ref{fig_SI_beta}B. There is no qualitative difference between these results and those of the A panels, indicating that the estimation performed by Eq.~(\ref{eq-noise_est_flat}) is not limited by the precision of the estimation of $b$ and $\alpha$ but instead by a different factor, such as the length of the dataset.

\subsection*{Posterior density and its maxima: experimental responses to natural stimuli}
\begin{figure}[h]
	\centering
	\includegraphics[width=0.8\paperwidth]{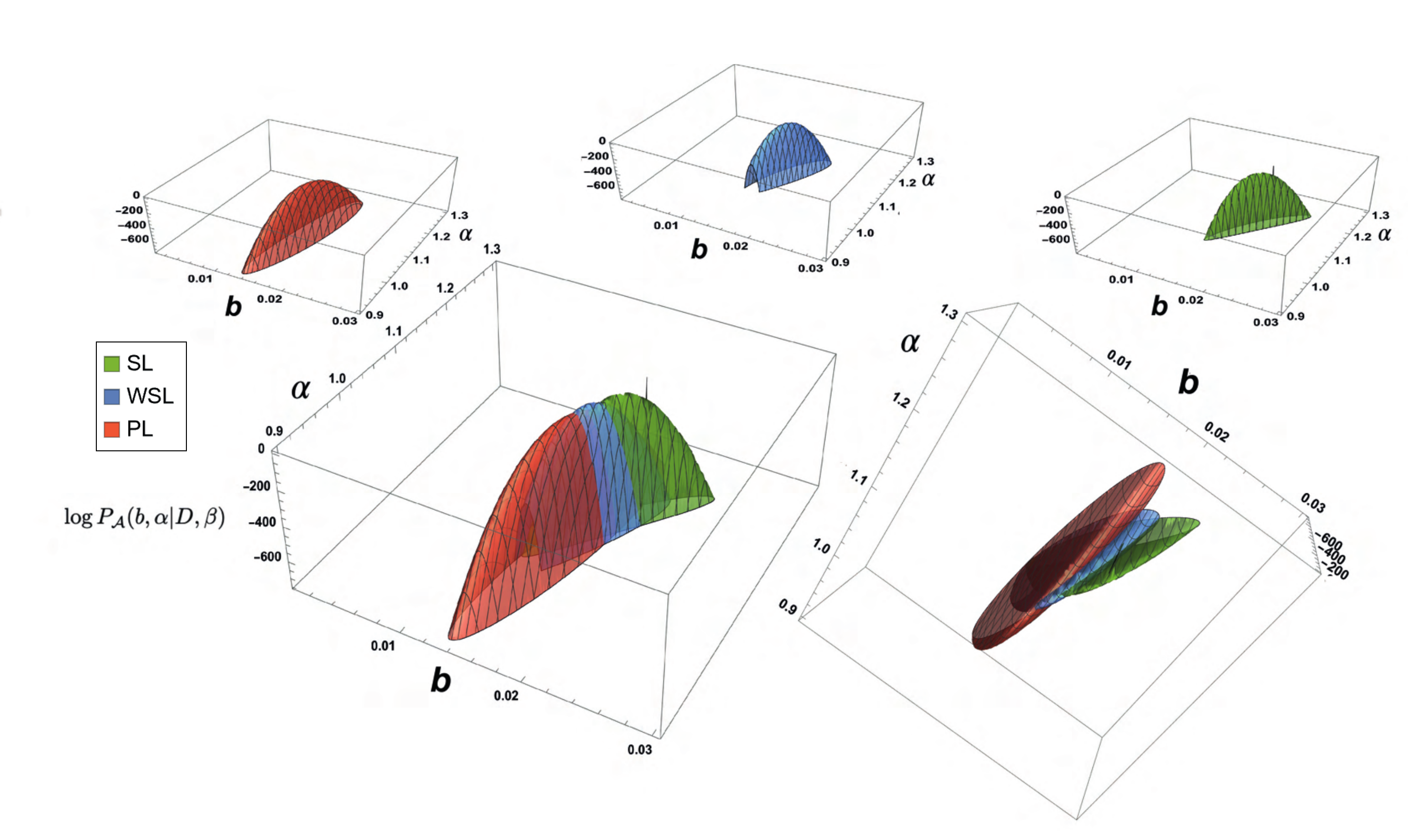}
	\caption{\textbf{Posterior densities for Recording 1.} Joint posterior density for parameters $b$ and $\alpha$ obtained for the first recording in the main dataset of~\cite{stringer2019high} for the SL model (green) and the WSL model (blue). For the PL model (red) we show the posterior density up to a constant scale factor. The top row shows each of these surfaces separately (in the order SL, WSL and PL, from left to right). The bottom row shows them combined. Only the points with indices $11$ to $500$ were used in the calculations, as was done in~\cite{stringer2019high}. A spike visible behind the SL surface and a missing piece of the WSL surface are only visual artefacts produced by the plotting software and don't reflect the true shape of those surfaces, which are smooth.}\label{fig_postdensr1}
\end{figure}

We now turn our attention to the results obtained for the dataset from~\cite{stringer2019high}. We calculate the posterior density for $b$ and $\alpha$ for each recording in that dataset using each of our models. For this purpose we use $x_{\rm min} = 11$ (the value used in~\cite{stringer2019high}) and a prior of the form~(\ref{eq-flat_prior}), with $\alpha_0 = 0, \alpha_M = 3, b_0 = 10^{-8}$ and $b_M = 0.05$. We expect this range to be appropriate for all recordings in this particular dataset, but we always inspect the posterior densities and the position of their maximum so that we can modify or extend the support of our prior if the maximum occurs at the border of this range. Both for the synthetic data as well as here, we perform our calculations using software (Wolfram Mathematica 12) which can directly manipulate the symbolic expressions presented earlier, and can apply built-in numerical methods to find extrema and calculate integrals. We verified the results obtained in this way by calculating the value of the posterior density for each point of a $10^3 \times 10^3$ grid spanning the range given by the prior, and simply finding the point on the grid for which the posterior is maximised; the results obtained with either method were consistent up to the minimum resolution provided by our calculations on the grid. Fig.~\ref{fig_postdensr1} shows the posterior densities obtained for the first recording in the dataset, using only the points with indices $11$ to $500$ as was done in~\cite{stringer2019high}. For the PL model, as a consequence of the practical difficulties we found when calculating the evidence, we show instead $\exp\left(\hat{\beta} \, (E^{\rm ML}_{\rm D}-E_{\rm D}(b,\alpha))\right)$, which is proportional to the posterior density for a flat prior. Each of the three surfaces presents a single maximum and then decays quickly, in a way that closely resembles a two-dimensional Gaussian. Those maxima are close to one another but do not coincide. The posterior densities for the other recordings in the dataset look very similar to these, as shown in Fig.~\ref{fig_SI_postden}.

It is important to note that if, for any reason, we were only interested in the best-fit value for just one of our parameters, typically $\alpha$, the correct density to look at would be the posterior density for that parameter only. We can obtain this density by marginalising the joint density for $b$ and $\alpha$ with respect to the parameter we are not interested in. 

In principle, obtaining the estimate for $\alpha$ from its marginal density, as opposed to from the joint maximisation of the density for $b$ and $\alpha$ could lead to quite dramatic differences. This is due to the fact that the $b$ and $\alpha$ are correlated, and the Gaussian approximation we used earlier to estimate our uncertainty in the best-fit values breaks down far away from the maximum. It is therefore important to investigate whether there is a substantial difference between the estimates found from marginalisation and joint maximisation for the dataset and models we analysed here. 

For the particular models and the dataset we are using, we found that the joint densities, shown in Figs.~\ref{fig_postdensr1} and~\ref{fig_SI_postden}, approximate a Gaussian very well around their maximum and fall off very quickly when moving away from it. Because of this, the marginal density is still well approximated by a Gaussian and the value of $\alpha$ at its maximum coincides with the one we obtained from the joint posterior density for $b$ and $\alpha$.

\begin{figure}[h!]
	\includegraphics[width=0.6\paperwidth]{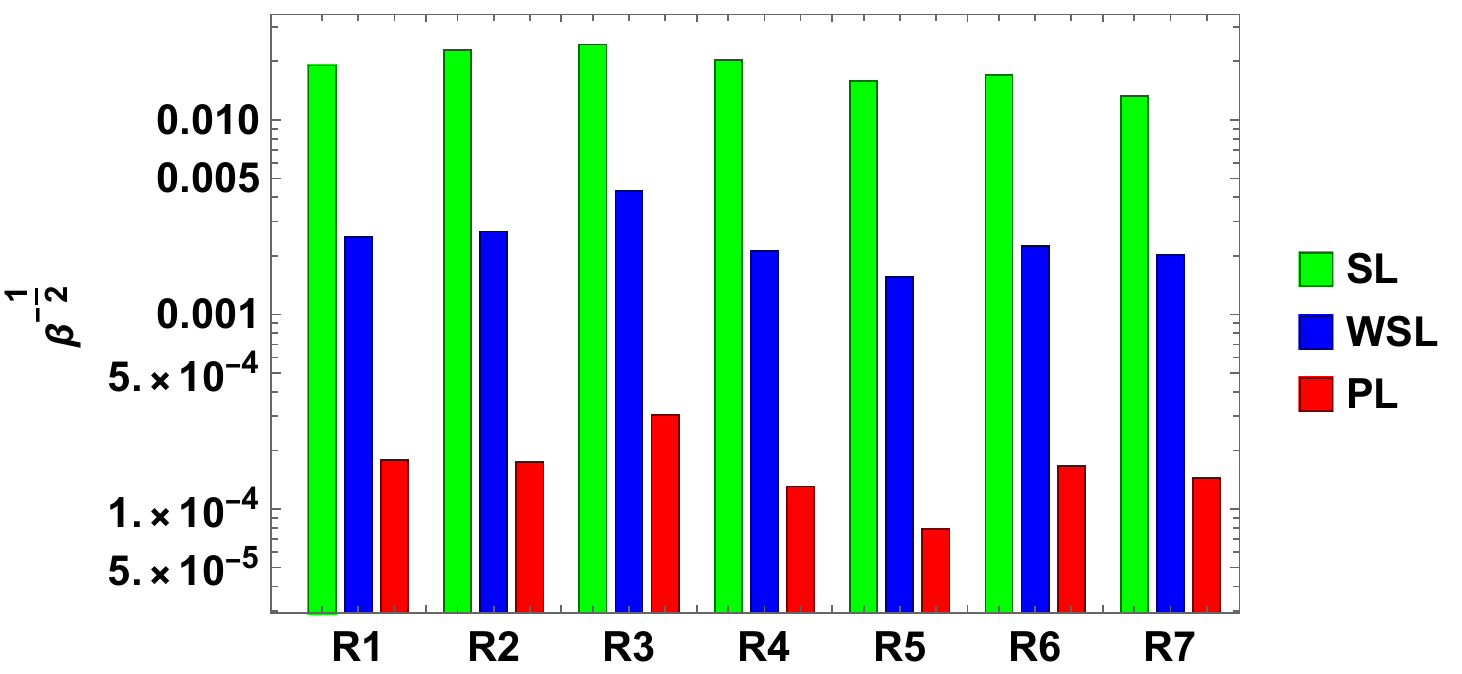}
	\caption{\textbf{Estimation of noise variance for the main experimental dataset.} Values of $\beta^{-1/2}$ obtained for each recording in the main dataset of~\cite{stringer2019high}, for each of the inference models, using Eq.~(\ref{eq-noise_est_flat}). When looking at this figure, it's important to remember that the definition of $\beta$ is different for each model (see \nameref{sec-model-desc}).}\label{fig_betas_barplot}
\end{figure}

When obtaining the posterior density, the Bayesian framework also provides us with an estimation of the inverse noise variance, $\beta$, for each of our models. Fig.~\ref{fig_betas_barplot} and Table~\ref{tab-betas} show the values of $\beta^{-\frac{1}{2}}$ obtained for each recording. When comparing these values, it is important to remember that the definition of $\beta$ is different in each model (see Eqs.~(\ref{eq-noisemodel}),~(\ref{eq-noise_2}) and definition of WSL). This leads, for example, to the values of $\beta^{-\frac{1}{2}}$ being an order of magnitude larger in the SL model than they are in the WSL model. We could also assign an uncertainty to our estimates of $\beta^{-\frac{1}{2}}$ by making use of the curvature of the evidence around its maximum; we omit such uncertainty calculations here. 

The full posterior distribution provides the most information about the result of our inference. From it, in addition to the maximum a posteriori (MAP) estimates of the parameters, we quantify the uncertainty of these estimates by calculating the covariance matrix for our parameters at the position of the maximum, as we did for the synthetic data. Noting that the vertical axis of Fig.~\ref{fig_postdensr1} is logarithmic, the figure reveals that these densities are indeed very sharp at the position of their maximum, implying a low uncertainty in the MAP estimates, as further described below.

\begin{figure}[h!]
    \begin{adjustwidth}{-0.4in}{0in}
	\includegraphics[width=\paperwidth]{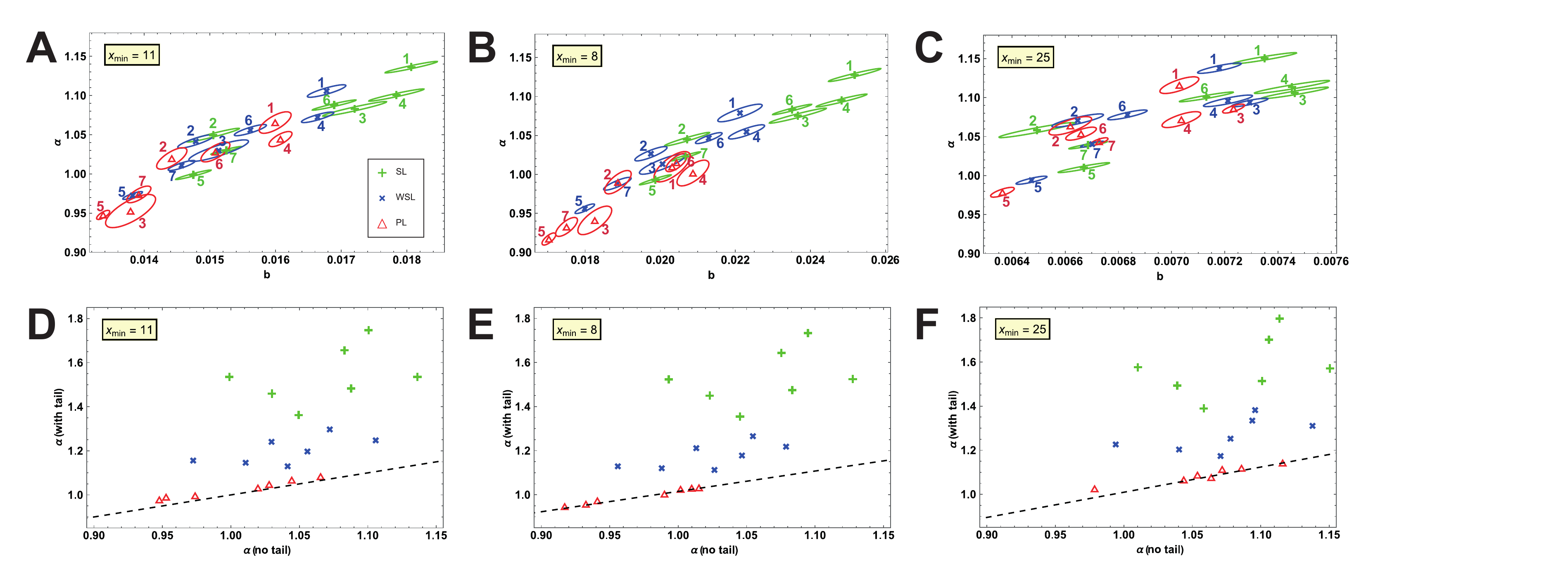}
	\caption{\textbf{Inference results for the main experimental dataset.} \textbf{(A-C)} Best-fit values for $b$ and $\alpha$ corresponding to the position of the maximum on their joint posterior distribution, for each recording in the main dataset of~\cite{stringer2019high}. The ellipse around each point indicates the $99\%$ confidence region. Green crosses and ellipses correspond to the SL model, blue $+$ symbols and ellipses to the WSL model and red triangles and ellipses to the PL model. The number next to each symbol indicates the recording to which it corresponds. Only the data points with indices $x_{\rm min}$ to $500$ of each recording were used in the calculations, with the value of $x_{\rm min}$ indicated in each panel. \textbf{(D-F)} Value of the $\alpha$-coordinate of the maximum of the joint posterior density when including or excluding the ``tail'' of the data from each recording (indices over $500$). The dashed line indicates the identity.} \label{fig_joint_bestfit}
	\end{adjustwidth}
\end{figure}

Fig.~\ref{fig_joint_bestfit}A shows the best-fit values for $b$ and $\alpha$ for each of the $7$ recordings in the dataset as estimated by each of the models. The ellipses around the points indicate the $99\%$ confidence region around the MAP estimates. The numerical values corresponding to this figure are provided in Table~\ref{S1_Table_500fitvalues}.     

We can learn from Fig.~\ref{fig_joint_bestfit}A that the best-fit values for each recording depend on the noise model. But are these differences important in any practical sense?

In~\cite{stringer2019high}, the authors used the experimental data to test their theoretical result, which states that for populations of neurons of size $N\to \infty$ an exponent of $\alpha>1$ is needed for differentiable codes. They thus performed an estimate of $\alpha$ for the 7 different recordings we have analysed here, finding that the average is slightly above $1$. Note that the WSL model is mathematically equivalent to the fitting procedure employed in~\cite{stringer2019high}, and we can indeed confirm that the average over recordings of the estimated value for the exponent, $\alpha$, using this model or the other two indeed yields values larger than one.

There are, however, a number of subtle observations to be made here. Firstly, one may ask if it makes sense to average over the values of $\alpha$ from different recordings. Indeed, Fig.~\ref{fig_synth_data} shows that when the data is generated using the same noise model as the one used for parameter estimation, this averaging makes sense. It could also makes sense in some other cases, such as when generating data with PL and inferring with one of the other models, since the results are mostly consistent across datasets. There are situations like the one when we generate with SL but infer with PL, where averaging does provide an estimate that is close to the true values even though the individual datasets are inconsistent with one another; however, in the absence of knowledge of how these data were generated and what the true values are, it would not make sense to average results that are so inconsistent with one another as these are. 
Secondly, we can see that when data is generated from PL but parameters are estimated using SL or WSL, the parameters tend to be overestimated and this could easily lead to concluding $\alpha>1$, when this is not so for the true values, as we have already explored in Figs.~\ref{fig_synth_datax5} and ~\ref{fig_SI_mockx10}. Although this may be a problem, depending on how large or small the overestimation is in practice, the following point is perhaps more curious: assuming each recording to be a sample from a similar power law whose exponent is to be estimated, Fig.~\ref{fig_joint_bestfit} does not seem to be qualitatively similar to the case when WSL is used on data generated from WSL (Fig.~\ref{fig_synth_data}E, blue crosses and ellipses). This suggests that the real data may not be consistent with the model used for estimation. In fact, the situation in Fig.~\ref{fig_joint_bestfit} does not seem to be similar to any of those in Fig.~\ref{fig_synth_data}, implying that the data is potentially not well-modeled by any of these three noise models, or perhaps not even by a power law to begin with. The latter point is of course one of model selection that we do not turn to, but already at this point the results beg the question of, given these discrepancies, how sure can one indeed be about the presence of a power law, let alone the value of its exponent. And this is all assuming that $x_{\rm min}$ and the cut-off values are known or knowledgeably chosen, issues that we turn to later on.

\subsection*{Posterior density and its maxima: experimental responses to low-dimensional stimuli}

As mentioned in the \nameref{sec-intro}, the theoretical analysis in \cite{stringer2019high} predicted that for a smooth population code the exponent of the power law should be larger than $1+2/d$. Besides the responses to natural stimuli that we analysed above, for which $d$ is supposedly very large and the lower bound on the exponent is around 1, the authors also studied recordings in response to stimuli sampled from spaces of lower dimension. 

\begin{figure}[h]
	\includegraphics[width=0.5\paperwidth]{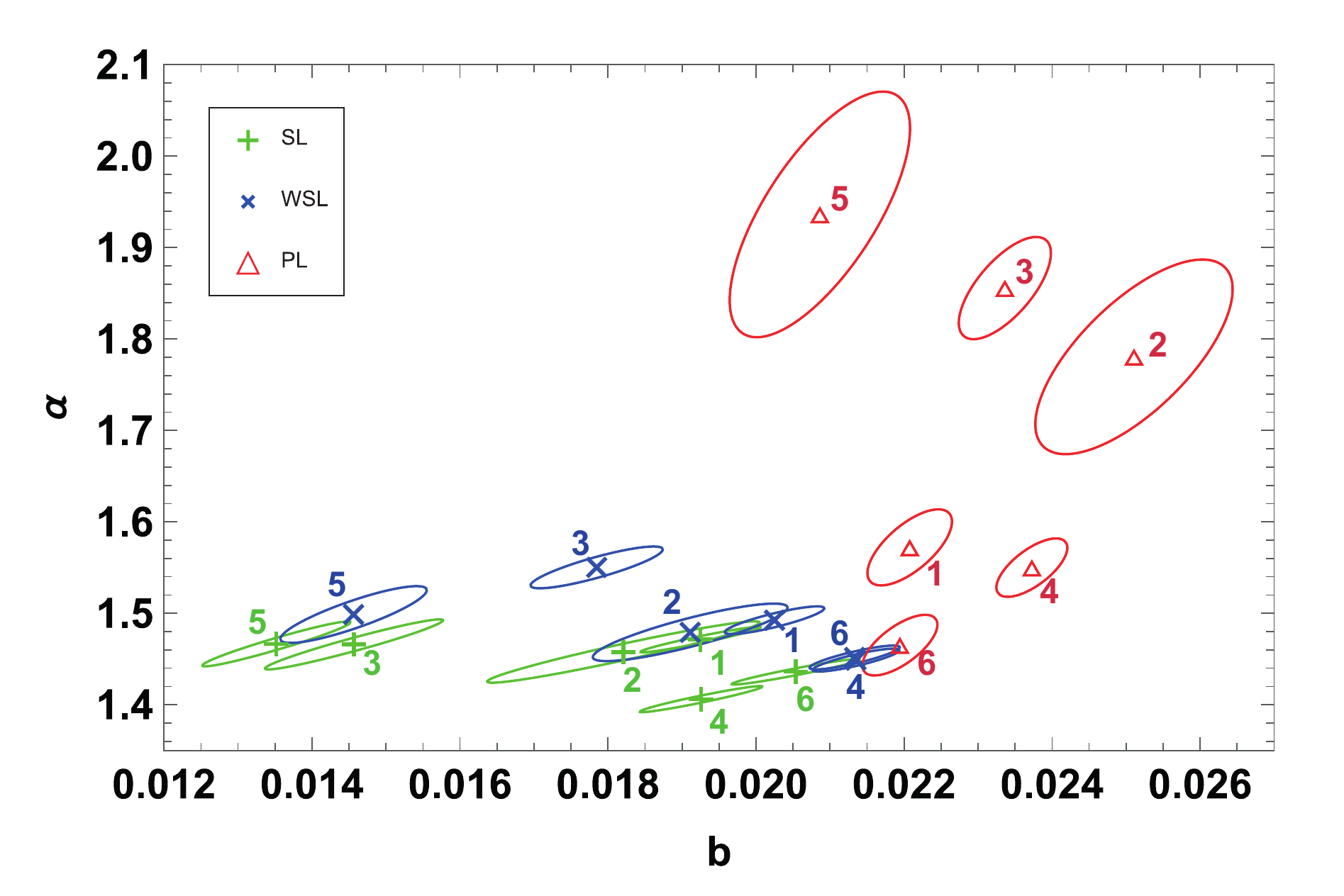}
	\caption{\textbf{Inference results for the 8D experimental dataset.} Best-fit values for $b$ and $\alpha$ corresponding to the position of the maximum on their joint posterior distribution, for each recording in the 8D dataset of~\cite{stringer2019high}. The ellipse around each point indicates the $99\%$ confidence region. Green crosses and lines correspond to the SL model, blue $+$ symbols and lines to the WSL model and red triangles and lines to the PL model. The number next to each symbol indicates the recording to which it corresponds. Only the data points with indices $11$ to $500$ of each recording were used in the calculations, as was done in~\cite{stringer2019high}.} \label{fig_8D}
\end{figure}

Here we also apply the Bayesian interpolation approach to the data recorded in response to 8-dimensional stimuli. For this dataset, the predicted lower bound on $\alpha$ would be 1.25. Fig.~\ref{fig_8D} shows the results of our estimations on this dataset. As opposed to the case of the responses to natural stimuli shown in Fig.~\ref{fig_joint_bestfit}, where for some recordings and inference models one could get estimates both below and above the predicted lower bound of 1, in the case of 8D stimuli the results are consistently above the lower bound of 1.25 with a reasonable margin. 

The 8D example thus portrays a case where, unlike that of natural stimuli, the dataset is sufficiently large and clear that, once a power law is assumed, the estimated exponent satisfies the bound and this is unlikely to depend on the noise model and other details.  

\subsection*{Exploring the effects of $x_{\rm min}$ and the upper cut-off}

Up to this point we have performed all our calculations using the same value of $x_{\rm min}$ as in~\cite{stringer2019high}, $x_{\rm min} = 11$. In principle, one can also include $x_{\rm min}$ as a parameter and find the best estimate for it. Here, in this section, we would like instead to simply explore how sensitive the estimates for $\alpha$ are to different choices for the value of $x_{\rm min}$. At the same time, the dataset from~\cite{stringer2019high} contains almost $3000$ data points for most of the recordings in the natural stimuli condition. However, the original authors use only the first $500$ points in their calculations. As was alluded to before, finding the optimal choices for each of these two parameters are themselves difficult estimation problems. Yet, being interested primarily in $\alpha$, we can study how these choices affect the results regarding $\alpha$ in synthetic and real data.

To explore the effect of including the ``tail'' cut off of the dataset (those points with indices larger than $500$), we repeat our calculations considering all datapoints available. The numerical results obtained for the best-fit values for $b$ and $\alpha$ in this case can be found in Table~\ref{S1_Table_fullfitvalues}. Figure~\ref{fig_joint_bestfit}D compares the $\alpha$-coordinate of the maximum of the joint posterior density obtained with and without the tail of the dataset with each model. We see clearly that, for this dataset, the PL model is far less sensitive than the other two to the inclusion of these additional datapoints, but all of them provide a larger value of $\alpha$ when all points are taken into account.

\begin{figure}[p]
	\includegraphics[width=0.7\paperwidth]{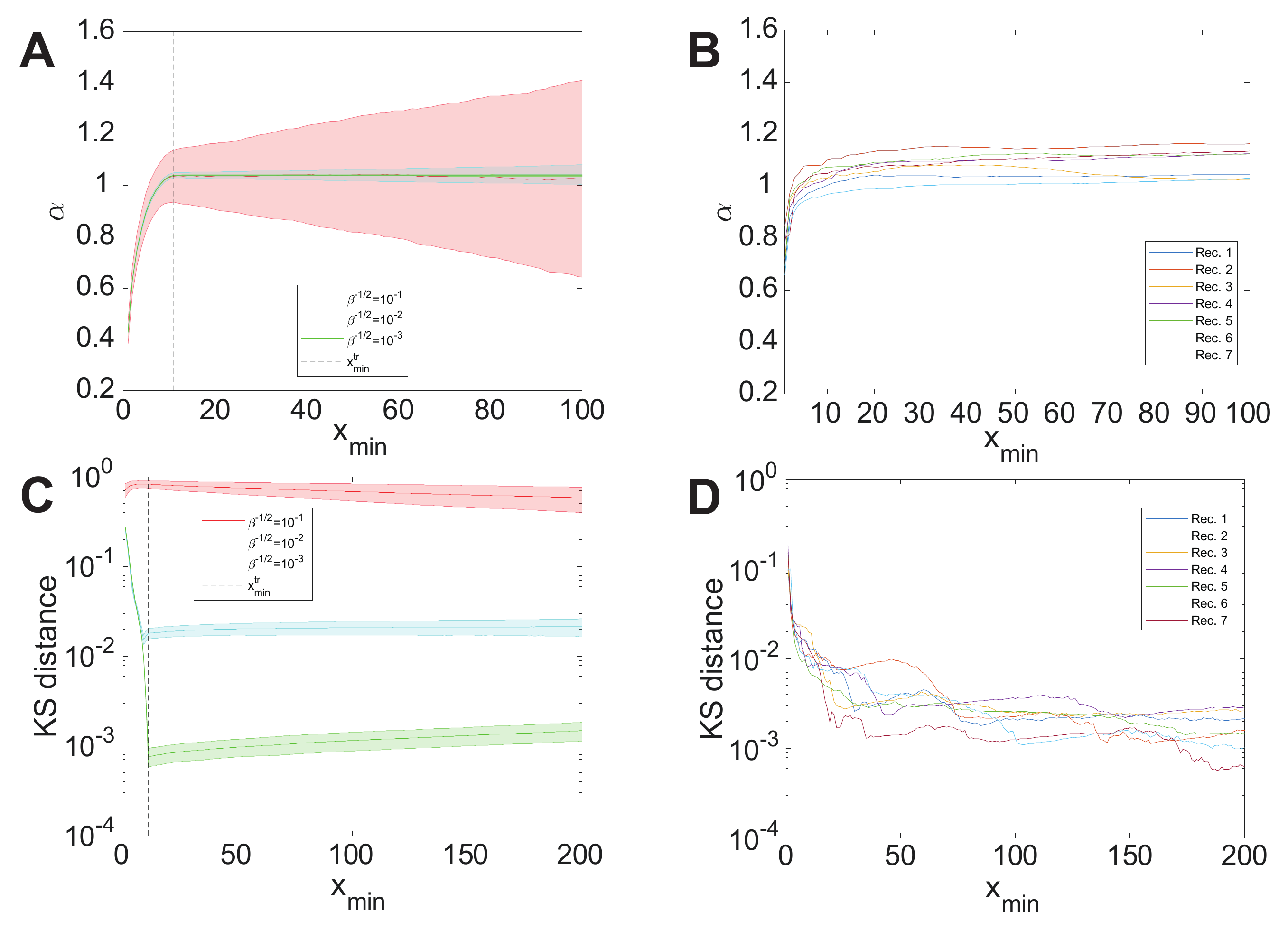}
	\caption{\textbf{Behaviour of $\alpha$ and the KS distance as we vary $x_{\rm min}$.} \textbf{(A)} Best estimate for $\alpha$ provided by least-squares averaged over $1000$ realizations of the noise, for each value of $\beta$, according to the WSL model on simulated data. Parameters are given in the main text. Shaded regions indicate $1$ standard deviation over the $1000$ realizations. The vertical dashed line indicates the true value of $x_{\rm min}$ used to generate the data. \textbf{(B)} Best estimate for $\alpha$ provided by least-squares for each recording in the main dataset of~\cite{stringer2019high} as we change the value of $x_{\rm min}$ used for inference. \textbf{(C)} Average Kolmogorov-Smirnov distance between the best-fit power law provided by least-squares and the data, as we vary the value of $x_{\rm min}$ used for inference. The average is over $1000$ realizations of the noise and each shaded region corresponds to $1$ standard deviation from that average. Different colours correspond to different values of $\beta$ as indicated. The vertical dashed line indicates the true value of $x_{\rm min}$ used to generate the data. \textbf{(D)} Kolmogorov-Smirnov distance between the best-fit power law provided by least-squares and the data from the main dataset in~\cite{stringer2019high}, as we vary the value of $x_{\rm min}$ used for inference. Each line corresponds to one of the $7$ recordings in the dataset, as indicated.} \label{fig_xmin_scan}
\end{figure}

Regarding the effect of $x_{\rm min}$ in our results, we can already see in Fig.~\ref{fig_joint_bestfit}B, C, E and F that changing this parameter can have dramatic consequences for our estimation of $\alpha$. To gain further insight regarding the effect of $x_{\rm min}$, we begin by studying a dataset with a known underlying value of $x_{\rm min}$ and the other parameters as we change the value of $x_{\rm min}$ used for inference. As before, we generate data with parameters $\alpha^{\rm tr} = 1.04$, $b^{\rm tr} = 0.01$ and $x^{\rm tr}_{\rm min} = 11$, according to Eq.~(\ref{eq-plaw}). Once again, For $x < x^{\rm tr}_{\rm min}$ we use the functional form $b \sqrt{x/x^{\rm tr}_{\rm min}}$. Regarding the noise, we use the WSL model so that comparisons to the results in~\cite{stringer2019high} can be made more directly. We add noise according to this model, and we explore different values for the variance parameter $\beta$; Fig.~\ref{fig_synth_data}B shows examples of what these datasets look like for a value of $\beta^{-\frac{1}{2}} = 10^{-3}$. We then find the best estimate for $\alpha$ by using the closed form least-squares solution implemented in the code provided by the authors of~\cite{stringer2019high}, which as we have said corresponds to the $\alpha$-coordinate of the maximum of the joint posterior distribution on $b$ and $\alpha$ when working with a flat prior. As discussed before, the marginal distribution for $\alpha$ would be more appropriate for this but the marginalisation procedure doesn't have a relevant effect in this case. Fig.~\ref{fig_xmin_scan}A shows what happens to the best-fit value of $\alpha$ as we change the value of $x_{\rm min}$ used in the inference process. Shaded areas indicate $1$ standard deviation over $1000$ realizations of the noise for each value of the noise variance. We see that the best-fit value of $\alpha$ changes rapidly until we reach the true value of $x_{\rm min}$ (the rate and direction of change will depend on the functional form we have chosen for $x < x^{\rm tr}_{\rm min}$) but the average value then stabilizes as soon as $x^{\rm tr}_{\rm min}$ is reached, as long as the noise variance is not too high. A few randomly chosen individual realizations of the noise can be seen in the Supporting Information, Fig.\ref{fig_si_indivnoise}.

Fig.~\ref{fig_xmin_scan}B shows the behavior of the best-fit estimate for $\alpha$ for each of the recordings in the main dataset from~\citen{stringer2019high} as we change the value of $x_{\rm min}$ used during the inference process (compare to Fig.\ref{fig_si_indivnoise}). We see that the value of $\alpha$ doesn't stabilize at $x_{\rm min} = 11$ in each recording but rather for larger values, if ever. This suggests that the value of $x_{\rm min}$ used in~\cite{stringer2019high} might have been underestimated. If so, that would imply that the best estimate values for $\alpha$ should have been higher than reported, though likely compatible with them within the level of uncertainty we reported earlier (assuming that the best estimate for $b$ moves in the same direction and by the right amount as well). The continuing drift of the best estimate for $\alpha$ towards higher values as $x_{\rm min}$ increases observed in some recordings suggests that the model used, either with regards to the power-law functional form or with regards to the noise model, might not be suitable for this dataset. Proper exploration of this question requires model selection.   

We further explore the effect of $x_{\rm min}$ by computing the Kolmogorov-Smirnov (KS) distance between the best-fit power law provided by WSL and the empirical distribution of values, as we change $x_{\rm min}$ \cite{clauset2009power}. Fig.~\ref{fig_xmin_scan}C shows the results for synthetic data. We see here that for low noise variance the KS distance descends rapidly until we reach the true value of $x_{\rm min}$, and then remains almost constant with very little variance around the average (over $1000$ trials). For low values of the noise variance and a functional form for $x < x^{\rm tr}_{\rm min}$ which is sufficiently different from the one after that point, identification of a good value for $x_{\rm min}$ seems plausible. Fig.~\ref{fig_xmin_scan}D also shows the KS distance as we change $x_{\rm min}$ but now using the dataset from~\cite{stringer2019high}. We see that for most recordings it's not easy to identify a clear minimum or transition point between two behaviors. This could once again be an indication of the model not being appropriate for the data, but proper model selection is required to provide some answer to that question.

\section{Conclusion}

Estimating the exponent of power laws has always been an important problem in the physics of critical phenomena and phase transitions, where power-law scalings are expected to appear at the point of transition and the exponents of the scaling laws are predicted to follow precise relationships with one another~\cite{stanley1999scaling}. It has thus always been an obsession of experimental physicists working on phase transition to measure the exponents precisely, with the same system being studied by different groups and different experimental methods; for example see table V of \cite{kadanoff1967static} summarising some of the experimental efforts to measure one critical exponent in magnetic transitions. As already noted in \cite{kadanoff1967static}, an important part of this endeavour has been to identify the critical region, that is, the range of parameters over which the data is to be fitted with a power law to estimate the exponent, and to study how the estimates depend on the choice of this region. In neuroscience, power-law relationships have also been reported, either via analysing theoretical models or by examining experimental data. With the ability to record from larger and larger populations of neurons, for longer periods of time and under varying conditions, in neuroscience too, power laws and their estimated exponents are playing a role. However, not enough attention has been paid to the caveats associated to drawing conclusions from data in this difficult inference task, in particular to the choice of the range of data to be analysed (which corresponds to the critical range in the theoretical physics alluded to above) or how much the amount of data and estimation method influence the results. 

The problem of asserting the presence of power laws by simply looking at the data in log-log scale and using linear regression has been noted by numerous people \cite{buchanan2008laws,stumpf2012critical}. As discussed before, some power laws, used either for describing probability distributions or as scaling laws, are predicated on the analysis of theoretical models \cite{fisher1967theory,bak1987self,stanley1999scaling}. Some of the studies that present experimental evidence for power laws rely not only on well-studied and sound theoretical models, but also on observations across a variety of settings and experimental measurement tools \cite{fisher1967theory,bak1987self,stanley1999scaling,kadanoff1967static}; such studies are not only limited to theoretical physics, but are also present in systems biology \cite{west1997general}. In other cases, for instance in many studies claiming the presence of power laws in network properties, results will largely go away when solid statistical models are used \cite{tanaka2005highly,stumpf2012critical,clauset2009power}. In neuroscience, however, it seems that power-law relationships are still declared only based on visual inspection, and exponents inferred using linear regression. Consequently, in this article we have proposed a reasonable first step for addressing some of the issues that arise in estimating the exponents of a power-law scaling: we perform inference through Bayesian interpolation, comparing the results from different noise models and looking at the consistency of the results. 

Given the task of inferring the values of the parameters for a power-law functional form interpolation, we asked ourselves how much the results of such inference would depend on our assumptions. To answer this question we began by recasting the problem in terms of the Bayesian framework of~\cite{MacKay92a}. We then proceeded to explore the effects of our assumptions both on synthetic data, for which we are sure about the presence of a power-law scaling and know the ground truth, and on recent experimental data from~\cite{stringer2019high}. We also observed that our choice of the value for the $x_{\rm min}$ parameter can have relevant consequences when inferring the values of our other parameters. A full Bayesian inference of $x_{\rm min}$ is desirable, but the methods we used here to explore the effects of this parameter on the inference process can also provide clues (or perhaps a prior) as to which values can be suitable choices for it.

Our results for the joint best-fit values of $b$ and $\alpha$ show such a degree of dependence on different choices, e.g. noise model, range of data or specific recording, that the estimated values are not compatible with one another. In fact, we found that for the responses to natural stimuli from~\cite{stringer2019high}, the same recording could in some cases be above or below the lower bound of 1 depending on the noise model employed during inference. However, the distribution of the best-fit parameters and their uncertainties in this case were quite different to what we observed in synthetic data, where the presence of a power law, its exponent and the noise model were known. This incompatibility suggests that either more data or a thorough process of model selection, perhaps aided by detailed modelling of the uncertainty and noise through the preprocessing of the data from measurement to interpolation, is required. For the case of responses to lower-dimensional stimuli, the results were more convincing: although the best estimates and uncertainties varied from recording to recording and depended on the noise model used for inference, they were all consistently above the lower bound suggested by the theory by a margin of $\sim15 \%$. 

Indeed, also in~\cite{stringer2019high} the difference between the estimated power-law exponent for low-dimensional stimuli and the corresponding lower bound was much larger than for the case of natural stimuli. The conclusion of~\cite{stringer2019high}, however, was that since in both cases their estimated exponents were still above the theoretical lower bound, the response manifolds for both natural and low-dimensional stimuli are smooth. As we discussed in this paper, one problem with their approach was that the exponents of different recordings were averaged. We argued that this is not justified. Furthermore, having quantified the uncertainties of the estimated exponents for each recording separately, and the dependence they have on choices such as the range of the data or the inference model, we show that it is not possible to conclusively state that the exponent in the case of natural scenes is above the lower bound and that the manifold of responses, as opposed to the case of low-dimensional stimuli, is smooth. 

The results can be interpreted in two ways: (a) that in the case of responses to high-dimensional stimuli the lower bound is in reality satisfied, as~\cite{stringer2019high} suggests, but that there is not enough data to show it conclusively or (b) that the manifold of responses of the mouse V1 to natural scenes is in reality not smooth, even if it is indeed smooth for lower-dimensional stimuli. Unfortunately, we cannot distinguish between these two possibilities given the data and tools we have available. However, it is important to note that the quantification of uncertainties, together with the realisation that the exponent in responses to natural stimuli (but not to lower-dimensional stimuli) could be above or below the bound depending on various factors, persuades one to seek an alternative interpretation of the data. It could indeed be the case that the stimulus-response mapping for low-dimensional stimuli is smooth, but for natural stimuli it hovers around a region where both smooth and non-smooth stimulus-response relationships are achievable, depending on different factors. For instance, attention is known to alter neural responses both at the single cell~\cite{reynolds2004attentional} and population levels~\cite{ruff2014attention,gomez2016neural}. It would thus be informative to study if the degree of smoothness of the stimulus-response relationship in a given recorded population is related to the degree of attention to the part of the scene that is retinotopically mapped to that population, and, from a broader perspective, other functional non-uniformities that such mapping entails ~\cite{sedigh2022retinotopy,heukamp2020topographic}.

Throughout this article we have made an effort to emphasise the importance of keeping our assumptions in mind and to consider the full posterior probability densities for our parameters when performing inference. The posterior density (whether joint or marginal, as appropriate) can provide much more information about the result of our inference than simply the position of its maximum. Crucially, it provides a way of quantifying the uncertainty in our estimations and the covariance or correlation of those estimates between the different parameters, which are always important to communicate. It is at the same time important to evaluate whether our distributions are well-approximated by a Gaussian around their maximum, as a failure to fulfil this condition could speak of a skewed inference where values away from the maximum in a certain direction are much more likely than in a different direction. 

Another crucial point to keep in mind is that, in the analysis that we performed here, the noise on each datapoint was considered to be independent. This assumption is unlikely to be true in real experimental data, as at least a part of the noise on each datapoint is due to the common recording techniques, similar pre-processing, etc. It would be important thus to also extend the Bayesian framework to models with correlated noise and evaluate the effects of such noise on any conclusions.

Finally, we reiterate that the results discussed here about the estimation of power-law exponents were made under the assumption that a power-law scaling is indeed present in the data. Arguing for the presence of this power-law scaling is a more difficult exercise, but the interpolation approach presented here provides a necessary first step towards this task of model selection.

\section*{Acknowledgments}
The authors are thankful to P.G.L. Porta Mana for insightful discussions and for pointing out Ref.~\cite{MacKay92a}, to Nicolai Waniek for valuable comments on an earlier version of the manuscript, and to Soledad Gonzalo Cogno for valuable discussions.

\bibliography{powerlaw.bib}

\newpage 

\section*{Supporting Information}\label{sec-supp}

\paragraph*{S1 Fig.}
{\bf Inference results on synthetic data with large noise.}
\renewcommand{\thefigure}{S1}

\begin{figure}[p]
	\centering
	\includegraphics[width=0.7\paperwidth]{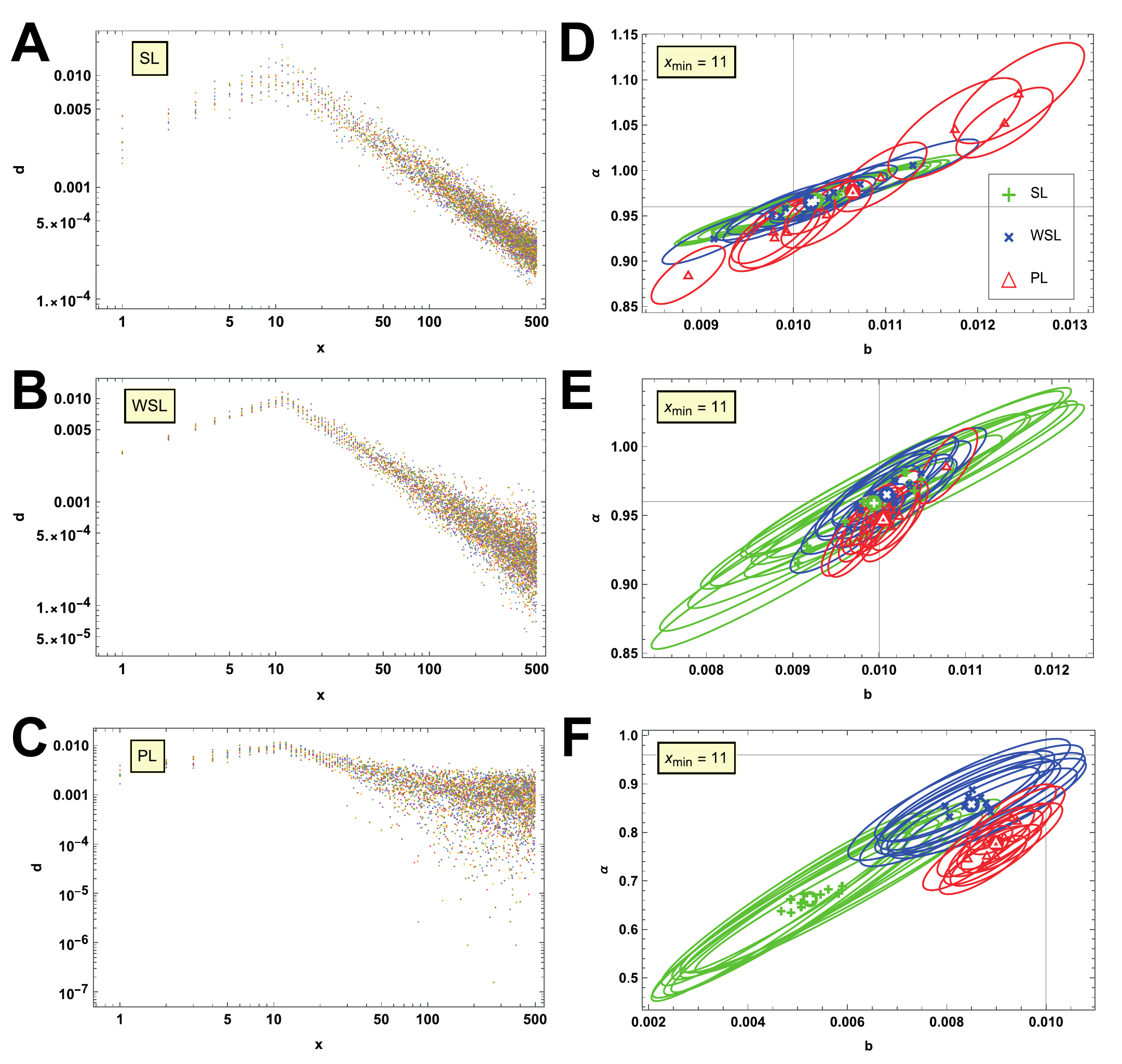}
	\caption{\textbf{Inference results on synthetic data with large noise and lower $\alpha$.} On the left column \textbf{(A-C)} we present $10$ datasets per panel, each generated according to one of our $3$ noise models, as indicated in the main text. Compared to Fig.~\ref{fig_synth_data}, the data presented here was created using a noise variance that is $100$ times larger and a lower value of the power-law exponent, $\alpha^{\rm tr} = 0.96$. Each colour corresponds to a different dataset. On the right column \textbf{(D-F)}, we summarize the results of performing inference on the datasets of the corresponding left panel with all $3$ models; the one that generated the data and the ones that didn't. We present this information in the form of the position of the maximum of the posterior distribution for $b$ and $\alpha$ as well as the $99\%$ confidence region ellipse around that maximum. To aid in the discussion, coloured disks with white symbols indicate the position of the average of the posterior maxima for the corresponding inference model, over the $10$ datasets. Black lines indicate the true values of the parameters. For the reasons explained in the text, the datasets used in \textbf{(F)} differ from those of \textbf{(C)} in the sign of some of the points with a large value of $x$.}\label{fig_SI_mockx10}
\end{figure}


\paragraph*{S2 Fig.}
{\bf Performance of the $\beta$ estimation with the SL and WSL models.}
\renewcommand{\thefigure}{S2}
\begin{figure}[p]
	\includegraphics[width=0.8\paperwidth]{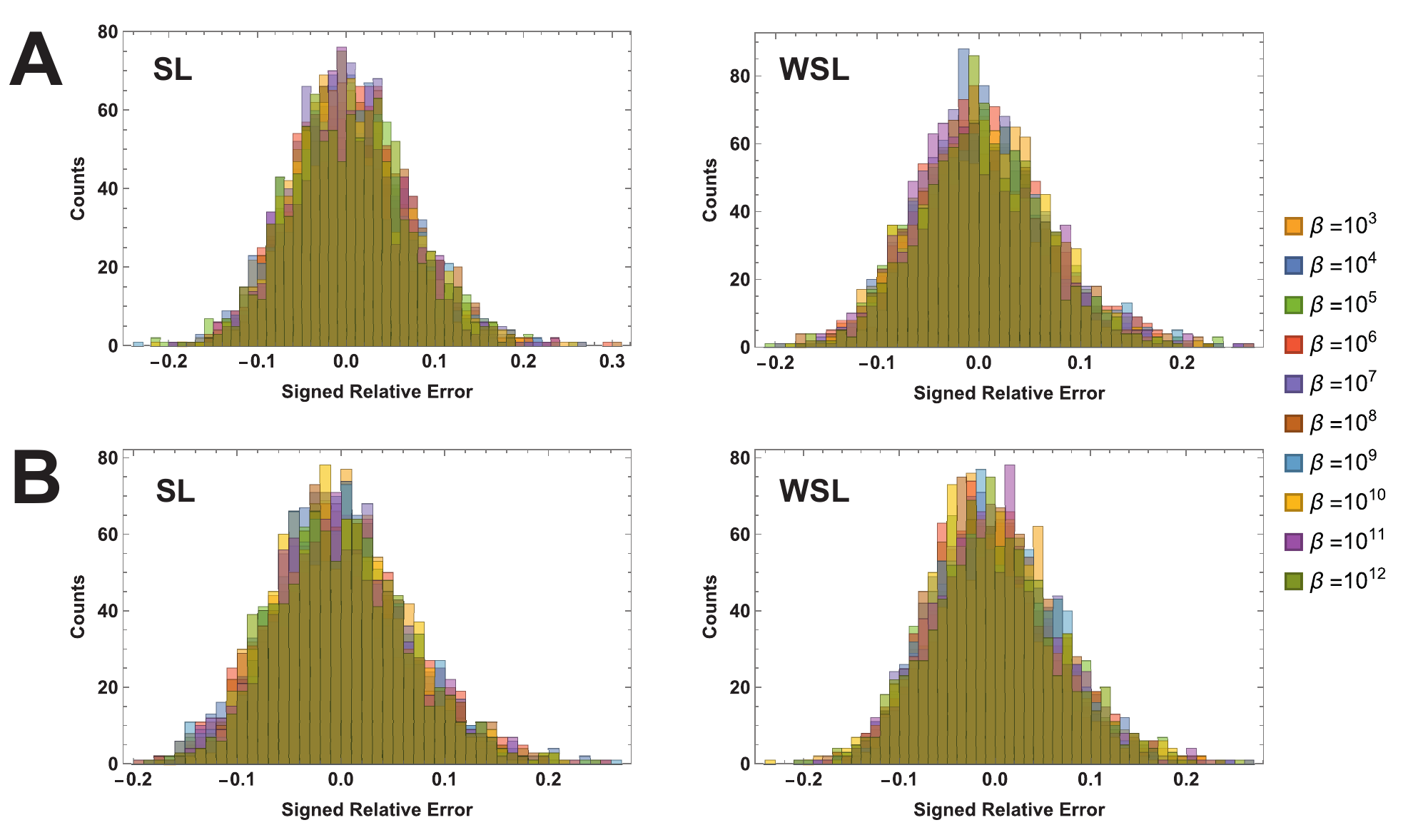}
	\caption{\textbf{Performance of the $\beta$ estimation with the SL and WSL models.} On the top row \textbf{(A)} we present histograms of the signed relative error in the estimation of $\beta$ when using Eq.~(\ref{eq-noise_est_flat}), after estimating $b$ and $\alpha$ with the SL and WSL models on corresponding synthetic datasets, and for a range of ground truth values of $\beta$ as indicated in the legend. Each histogram was built from $1000$ datasets per ground truth value of $\beta$. On the bottom row \textbf{(B)}, we show the same type of histograms, but now the ground truth value of $b$ and $\alpha$ has been employed when using Eq.~(\ref{eq-noise_est_flat}). Bin widths are $0.01$ for all panels.}\label{fig_SI_beta}
\end{figure}


\paragraph*{S3 Fig.}
{\bf Posterior density surfaces for recordings $2$ to $7$.}
\renewcommand{\thefigure}{S3}
\begin{figure}[p]
	\includegraphics[width=0.85\paperwidth]{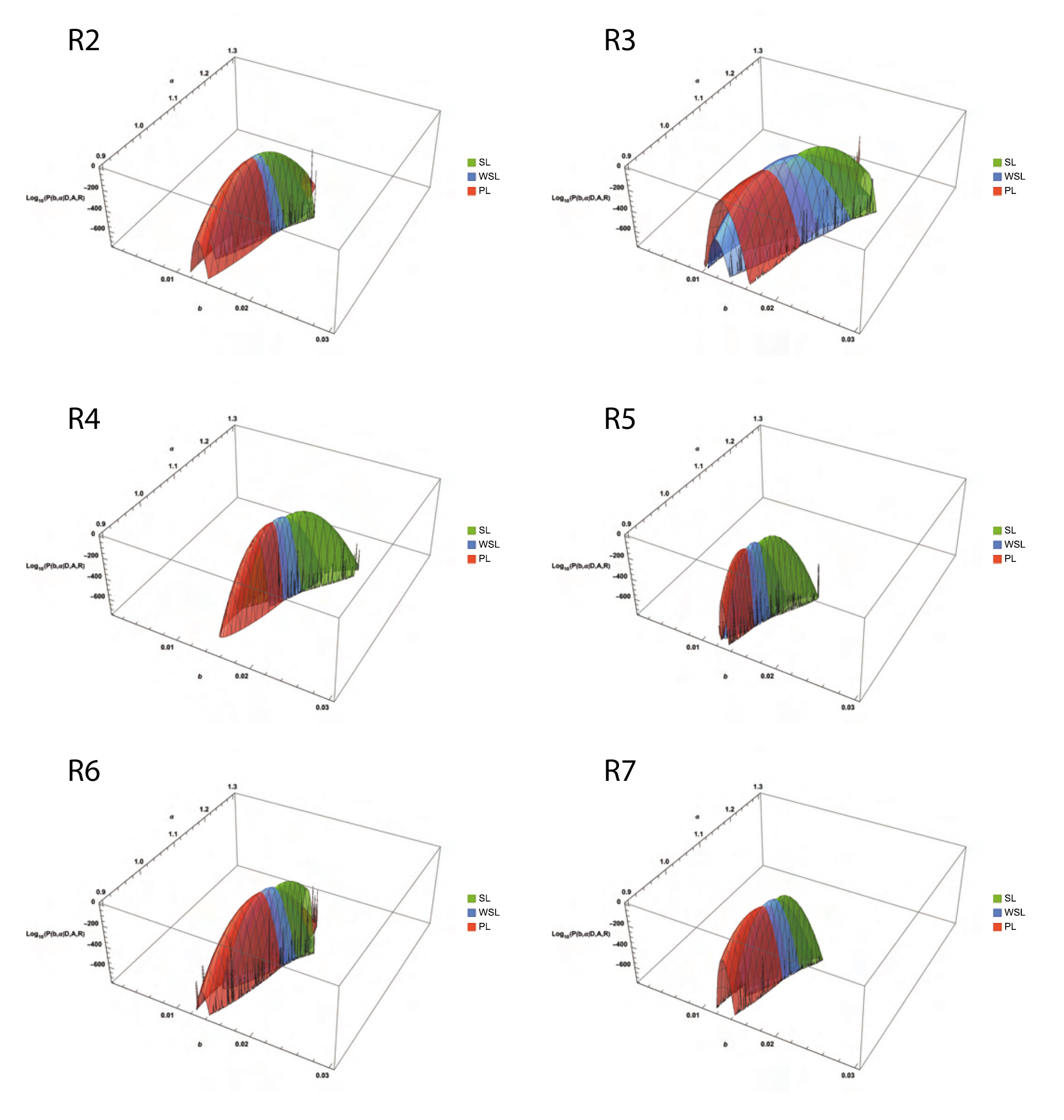}
	\caption{\textbf{Posterior density surfaces for recordings $2$ to $7$.} Joint posterior density for parameters $b$ and $\alpha$ obtained for recordings number $2$ to $7$ in the main dataset of~\cite{stringer2019high} for the SL (green) and the WSL models (blue). For the PL model (red) we show the posterior density up to a constant scale factor. Only the points with indices $11$ to $500$ were used in the calculations, as was done in~\cite{stringer2019high}. The spikes and missing parts visible on the edges of some of the surfaces are only visual artefacts produced by the plotting software and don't reflect their true shapes, which are smooth.}\label{fig_SI_postden}
\end{figure}


\paragraph*{S4 Fig.}
{\bf Behaviour of $\alpha$ on individual realizations of the noise for synthetic data as we vary $x_{\rm min}$.}
\renewcommand{\thefigure}{S4}
\begin{figure}[p]
	\includegraphics[width=0.9\textwidth]{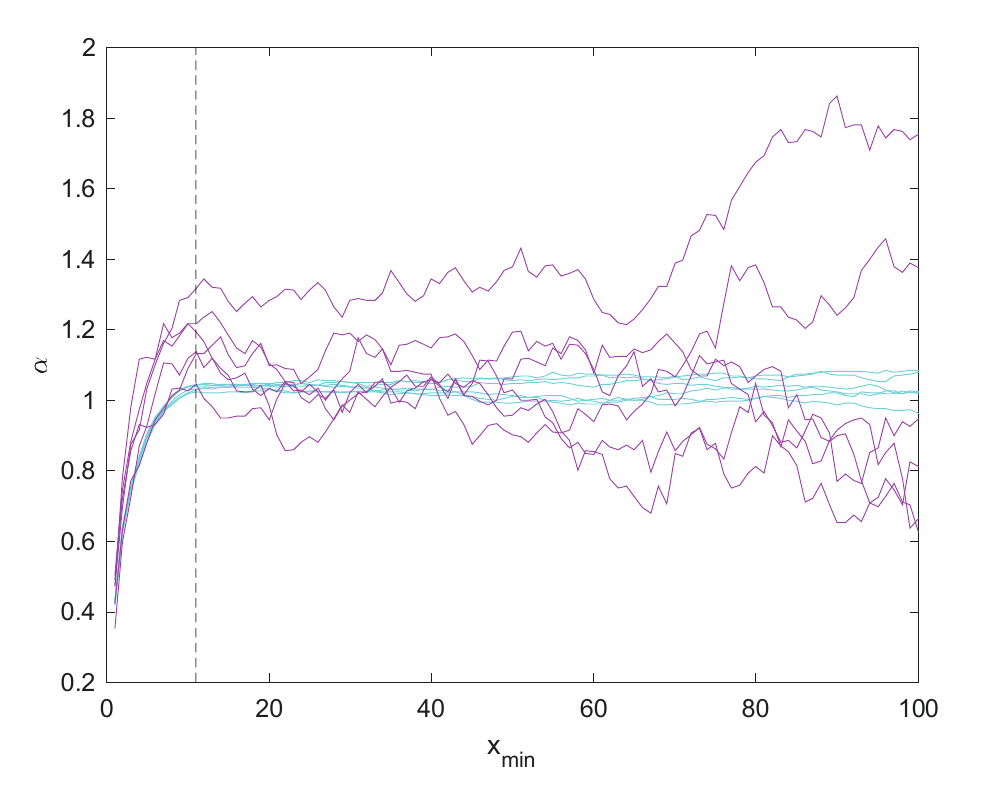}
	\caption{\textbf{Behaviour of $\alpha$ on individual realizations of the noise for synthetic data as we vary $x_{\rm min}$.} Best estimate for $\alpha$ provided by least-squares for individual realizations of the noise according to the WSL model on simulated data. Parameters are given in the main text. Cyan lines correspond to $\beta^{-\frac{1}{2}} = 10^{-2}$ and magenta lines correspond to $\beta^{-\frac{1}{2}} = 10^{-1}$. The vertical dashed line indicates the true value of $x_{\rm min}$ used for generating the data.}\label{fig_si_indivnoise}
\end{figure}

\newpage

\paragraph*{S1 Table. Values of $\beta^{-\frac{1}{2}}$ corresponding to Fig.\ref{fig_betas_barplot}}

\renewcommand{\thetable}{S1}
\begin{table}[!ht]
		\centering
		\caption{
			{\bf Values of $\beta^{-\frac{1}{2}}$ obtained for each recording and each model.}}
	\begin{tabular}{|c||c|c|c|c|c|c|c|} \hline
		 $\beta^{-\frac{1}{2}}$ & R1 & R2 &  R3 & R4 & R5 & R6 & R7 \\ \hline \hline
	SL & $1.9 \times 10^{-2}$ & $2.3 \times 10^{-2}$ & $2.4 \times 10^{-2}$ & $2.0 \times 10^{-2}$ & $1.6 \times 10^{-2}$ & $1.7 \times 10^{-2}$ & $1.3 \times 10^{-2}$ \\ \hline
	WSL & $2.5 \times 10^{-3}$ & $2.7 \times 10^{-3}$ & $4.3 \times 10^{-3}$ & $2.1 \times 10^{-3}$ & $1.6 \times 10^{-3}$ & $2.2 \times 10^{-3}$ & $2 \times 10^{-3}$ \\ \hline
	PL & $1.8 \times 10^{-4}$ & $1.7 \times 10^{-4}$ & $3.1 \times 10^{-4}$ & $1.3 \times 10^{-4}$ & $8 \times 10^{-5}$ & $1.7 \times 10^{-4}$ & $1.5 \times 10^{-4}$ \\ \hline	
	\end{tabular}
	\label{tab-betas}
\end{table} 


\paragraph*{S2 Table. Best-fit values corresponding to Fig.\ref{fig_joint_bestfit}.}

\renewcommand{\thetable}{S2}
\begin{table}[!ht]
	\centering
	\caption{
		{\bf Best-fit values corresponding to Fig.\ref{fig_joint_bestfit}.} Best-fit values for $b$ and $\alpha$ according to each of the $3$ models we consider, using only data in the range used in~\cite{stringer2019high}, corresponding to $x_{\rm min} = 11$ and $x_{\rm max} = 500$. Uncertainty is indicated in parenthesis for the last significant digit and corresponds to a standard deviation in the direction given by that parameter. For compactness, correlation matrices are provided instead of covariance matrices; they should be completed by symmetry.}
	\begin{tabular}{|c|c|c|cc|c|cc|c|cc|} \cline{3-11}
		\multicolumn{2}{c|}{\multirow{2}{*}{}} & \multicolumn{3}{c|}{SL}  &  \multicolumn{3}{c|}{WSL}  & \multicolumn{3}{c|}{PL}  \\ \cline{3-11} 
		\multicolumn{2}{c|}{} & Fit & \multicolumn{2}{c|}{Correlations} & Fit & \multicolumn{2}{c|}{Correlations} & Fit & \multicolumn{2}{c|}{Correlations} \\ \cline{3-11} \hline 
		\multirow{2}{*}{R1} & $b$ & 0.0181(1) & 1 &  & 0.0168(1) & 1 & & 0.01599(8) & 1 &  \\ 
		& $\alpha$ & 1.137(2) & 0.96 & 1 & 1.107(3) & 0.86 & 1 & 1.065(5) & 0.71 & 1 \\ \hline
		\multirow{2}{*}{R2} & $b$ & 0.0151(1) & 1 &  & 0.01480(9) & 1 & & 0.01441(8) & 1 &  \\ 
		& $\alpha$ & 1.050(3) & 0.96 & 1 & 1.041(3) & 0.86 & 1 & 1.020(7) & 0.72 & 1 \\ \hline
		\multirow{2}{*}{R3} & $b$ & 0.0172(2) & 1 &  & 0.0151(1) & 1 & & 0.0138(1) & 1 &  \\ 
		& $\alpha$ & 1.083(3) & 0.96 & 1 & 1.029(5) & 0.86 & 1 & 0.954(7) & 0.72 & 1 \\ \hline
		\multirow{2}{*}{R4} & $b$ & 0.0179(1) & 1 &  & 0.01660(8) & 1 & & 0.01607(6) & 1 &  \\ 
		& $\alpha$ & 1.101(3) & 0.96 & 1 & 1.071(2) & 0.86 & 1 & 1.044(3) & 0.72 & 1 \\ \hline
		\multirow{2}{*}{R5} & $b$ & 0.01480(9) & 1 &  & 0.01380(5) & 1 & & 0.01337(3) & 1 &  \\ 
		& $\alpha$ & 0.999(2) & 0.96 & 1 & 0.972(2) & 0.86 & 1 & 0.948(2) & 0.73 & 1 \\ \hline
		\multirow{2}{*}{R6} & $b$ & 0.0170(1) & 1 &  & 0.01560(8) & 1 & & 0.01508(7) & 1 &  \\ 
		& $\alpha$ & 1.089(2) & 0.96 & 1 & 1.056(2) & 0.86 & 1 & 1.029(4) & 0.72 & 1 \\ \hline
		\multirow{2}{*}{R7} & $b$ & 0.01520(8) & 1 &  & 0.01460(7) & 1 & & 0.01392(6) & 1 &  \\ 
		& $\alpha$ & 1.029(2) & 0.96 & 1 & 1.011(2) & 0.86 & 1 & 0.975(4) & 0.73 & 1 \\ \hline	
	\end{tabular} \label{S1_Table_500fitvalues}
\end{table}

\paragraph*{S3 Table.  Best-fit values when using the full tail of the dataset.}

\renewcommand{\thetable}{S3}
\begin{table}[!h]
		\centering
		\caption{
			{\bf Best-fit values when using the full tail of the dataset.} Best-fit values for $b$ and $\alpha$ according to each of the $3$ models we consider, using all datapoints available after $x_{\rm min} = 11$. Uncertainty is indicated in parenthesis for the last significant digit and corresponds to a standard deviation in the direction given by that parameter. For compactness, correlation matrices are provided instead of covariance matrices; they should be completed by symmetry. Comparing with Table~\ref{S1_Table_500fitvalues} we see that the PL model is less sensitive to the inclusion of this additional data for this particular dataset.}
		\begin{tabular}{|c|c|c|cc|c|cc|c|cc|} \cline{3-11}
			\multicolumn{2}{c|}{\multirow{2}{*}{}} & \multicolumn{3}{c|}{SL}  &  \multicolumn{3}{c|}{WSL}  & \multicolumn{3}{c|}{PL}  \\ \cline{3-11} 
			\multicolumn{2}{c|}{} & Fit & \multicolumn{2}{c|}{Correlations} & Fit & \multicolumn{2}{c|}{Correlations} & Fit & \multicolumn{2}{c|}{Correlations} \\ \cline{3-11} \hline 
			\multirow{2}{*}{R1} & $b$ & 0.063(2) & 1 &  & 0.0210(2) & 1 & & 0.01617(4) & 1 &  \\ 
			& $\alpha$ & 1.535(8) & 0.98 & 1 & 1.248(4) & 0.86 & 1 & 1.084(2) & 0.68 & 1 \\ \hline
			\multirow{2}{*}{R2} & $b$ & 0.042(1) & 1 &  & 0.0172(2) & 1 & & 0.01454(3) & 1 &  \\ 
			& $\alpha$ & 1.362(7) & 0.98 & 1 & 1.130(3) & 0.86 & 1 & 1.032(2) & 0.69 & 1 \\ \hline
			\multirow{2}{*}{R3} & $b$ & 0.108(5) & 1 &  & 0.0213(3) & 1 & & 0.01416(6) & 1 &  \\ 
			& $\alpha$ & 1.656(9) & 0.98 & 1 & 1.241(4) & 0.86 & 1 & 0.990(3) & 0.68 & 1 \\ \hline
			\multirow{2}{*}{R4} & $b$ & 0.141(7) & 1 &  & 0.0240(4) & 1 & & 0.01632(3) & 1 &  \\ 
			& $\alpha$ & 1.74(1) & 0.98 & 1 & 1.297(5) & 0.86 & 1 & 1.068(2) & 0.69 & 1 \\ \hline
			\multirow{2}{*}{R5} & $b$ & 0.083(3) & 1 &  & 0.0186(2) & 1 & & 0.01367(3) & 1 &  \\ 
			& $\alpha$ & 1.535(9) & 0.98 & 1 & 1.156(4) & 0.86 & 1 & 0.978(1) & 0.69 & 1 \\ \hline
			\multirow{2}{*}{R6} & $b$ & 0.060(2) & 1 &  & 0.0196(2) & 1 & & 0.01529(3) & 1 &  \\ 
			& $\alpha$ & 1.483(7) & 0.98 & 1 & 1.197(3) & 0.86 & 1 & 1.048(2) & 0.69 & 1 \\ \hline
			\multirow{2}{*}{R7} & $b$ & 0.063(2) & 1 &  & 0.0182(2) & 1 & & 0.01414(3) & 1 &  \\ 
			& $\alpha$ & 1.460(8) & 0.98 & 1 & 1.146(3) & 0.86 & 1 & 0.997(2) & 0.69 & 1 \\ \hline	
		\end{tabular} \label{S1_Table_fullfitvalues}
\end{table}

\nolinenumbers

\end{document}